\documentclass[useAMS,usenatbib]{mn2e}

\usepackage{epsfig}
\usepackage{amssymb,amsmath}
\usepackage{graphicx}
\usepackage{natbib}
\usepackage{rotating}
\usepackage{subfigure}
\usepackage{verbatim}
\usepackage{color}
\usepackage{multirow}
\usepackage{fancyhdr}
\usepackage{mathptmx}

\title[The ``Count Matching'' Approach for SMGs in the ECDF-S]{Properties of Submillimeter Galaxies in a Semi-analytic Model using the ``Count Matching'' Approach: Application to the ECDF-S}
\author[A. M. Mu\~noz Arancibia et al.]{Alejandra M. Mu\~noz Arancibia$^{1}$\thanks{E-mail: amunoz@astro.puc.cl}, Felipe P. Navarrete$^{2,3}$, Nelson D. Padilla$^{1,4}$, \newauthor Sof\'ia A. Cora$^{5,6}$, Eric Gawiser$^{7}$, Peter Kurczynski$^{7}$ and Andr\'es N. Ruiz$^{6,8,9}$\\
$^{1}$ Instituto de Astrof\'isica, Pontificia Universidad Cat\'olica de Chile, Av. Vicu\~na Mackenna 4860, Santiago, Chile\\
$^{2}$ Argelander-Institut f\"ur Astronomie, Auf dem H\"ugel 71, Bonn, D-53121, Germany\\
$^{3}$ Max-Planck-Institut f\"ur Radioastronomie, Auf dem H\"ugel 69, Bonn, D-53121, Germany\\
$^{4}$ Centro de Astro-Ingenier\'ia, Pontificia Universidad Cat\'olica de Chile, Av. Vicu\~na Mackenna 4860, Santiago, Chile\\
$^{5}$ Instituto de Astrof\'isica de La Plata (CCT La Plata, CONICET, UNLP), Facultad de Ciencias Astron\'omicas y Geof\'isicas,\\Universidad Nacional de La Plata, Paseo del Bosque s/n, B1900FWA, La Plata, Argentina\\
$^{6}$ Consejo Nacional de Investigaciones Cient\'ificas y T\'ecnicas, Rivadavia 1917, C1033AAJ Buenos Aires, Argentina\\
$^{7}$ Department of Physics and Astronomy, Rutgers, The State University of New Jersey, Piscataway, NJ 08854, USA\\
$^{8}$ Instituto de Astronom\'ia Te\'orica y Experimental (IATE, CCT C\'ordoba, CONICET, UNC), Laprida 854, X5000BGR, C\'ordoba, Argentina\\
$^{9}$ Observatorio Astron\'omico de C\'ordoba, Universidad Nacional de C\'ordoba, Laprida 854, C\'ordoba, X5000GBR, Argentina}

\date{Accepted 2014 October 23. Received 2014 October 16; in original form 2014 April 19}

\pagerange{\pageref{firstpage}--\pageref{lastpage}} \pubyear{2014}

\begin{document}

\label{firstpage}

\maketitle

\begin{abstract}
We present a new technique for modeling submillimeter galaxies (SMGs): the ``Count Matching'' approach. Using lightcones drawn from a semi-analytic model of galaxy formation, we choose physical galaxy properties given by the model as proxies for their submillimeter luminosities, assuming a monotonic relationship. As recent interferometric observations of the Extended Chandra Deep Field South show that the brightest sources detected by single-dish telescopes are comprised by emission from multiple fainter sources, we assign the submillimeter fluxes so that the combined LABOCA plus bright-end ALMA observed number counts for this field are reproduced. After turning the model catalogs given by the proxies into submillimeter maps, we perform a source extraction to include the effects of the observational process on the recovered counts and galaxy properties. We find that for all proxies, there are lines of sight giving counts consistent with those derived from LABOCA observations, even for input sources with randomized positions in the simulated map. Comparing the recovered redshift, stellar mass and host halo mass distributions for model SMGs with observational data, we find that the best among the proposed proxies is that in which the submillimeter luminosity increases monotonically with the product between dust mass and SFR. This proxy naturally reproduces a positive trend between SFR and bolometric IR luminosity. The majority of components of blended sources are spatially unassociated.
\end{abstract}


\section{Introduction}\label{intro}

Fitting submillimeter galaxies (SMGs) into the current theory of galaxy formation has been a challenge since their discovery. They are the most luminous star-forming sources at the epoch where star formation peaks, being detected by their redshifted far infrared (FIR) emission from cold dust in the submillimeter (submm) wavebands. SMGs were detected for the very first time in the Hubble Deep Field at $850\,\mu\textmd{m}$ (\citealt{Hughes1998}, \citealt{Barger1998}) using the Submillimeter Common-User Bolometer Array on the James Clerk Maxwell Telescope (JCMT/SCUBA, \citealt{Holland1999}). The advantage of observing SMGs is their negative $k$-correction: the observed flux coming from star-forming regions remains nearly constant with redshift at these wavelengths, so they can be detected out to $z\sim10$ \citep{Blain2002}.

Many observational efforts have been done with the attempt of mapping the global cosmic star formation history and the build-up of the overall stellar mass content of the Universe through SMGs, as they are useful tracers of the obscured star formation at high redshifts, complementing the cosmic census carried on at shorter wavelengths \citep{Blain1999b}. These include surveys using single-dish telescopes (\citealt{Coppin2006}, \citealt{Weiss2009}, \citealt{Chen2013}, just to name a few) and interferometric observations, both in the continuum and targeting emission lines (\citealt{Tacconi2006}, \citealt{Smolcic2012}, \citealt{Weiss2013}, \citealt{Hodge2013}, \citealt{Hatsukade2013}, \citealt{Chen2014}, among others). 

Identification of counterparts at other wavelengths for different samples of SMGs has been carried out, either through deep optical/near infrared (NIR) spectroscopy or deep intermediate/high resolution radio imaging. This allows the spectroscopic follow-up of SMGs, as well as multiwavelength studies. The largest sample of SMGs with spectroscopic follow-up to date was obtained by \citet{Chapman2005}, who measured a median spectroscopic redshift of 2.2 for 73 SMGs belonging to several fields, having a median $850\,\mu\textmd{m}$ flux density of $5.7\,\textmd{mJy}$. Moreover, using multiwavelength data \citet{Wardlow2011} derived a median photometric redshift of $2.2\pm0.1$ for 78 single-dish detected SMGs in the Extended Chandra Deep Field South (ECDF-S) over $\sim4\,\textmd{mJy}$ at $870\,\mu\textmd{m}$. More recent studies have been successful identifying counterparts of single-dish detected SMGs using interferometers at millimeter (mm) wavebands. For instance, \citet{Smolcic2012} performed interferometric observations at $1.3\,\textmd{mm}$ of SMGs in the central region of the Cosmic Evolution Survey (COSMOS) field, which had been discovered through single-dish observations at $870\,\mu\textmd{m}$ down to a signal-to-noise ratio (S/N) of 3.8; complementing this with additional multiwavelength catalogs, they derived a mean photometric redshift of $2.6\pm0.4$ for 16 of these SMGs.

Having redshift information and multiwavelength photometry allows the performance of spectral energy distribution (SED) fitting, and then deriving other SMG properties as stellar mass and star formation rate (SFR). From several studies exploring these properties (\citealt{Smail2004}, \citealt{Borys2005}, \citealt{Michalowski2010}, \citealt{Wardlow2011} and others), there is now some consensus regarding the quite high SFRs of single-dish detected SMGs, having several hundreds and even thousands of $\textmd{M}_{\odot}/\textmd{yr}$ and hosting significant stellar populations. These findings place them among the most powerful starburst galaxies in the Universe. Additionally, making these sources evolve through models, it has been found that SMGs are likely the progenitors of local luminous early-type galaxies (\citealt{Smail2004}, \citealt{Swinbank2006} and \citealt{Wardlow2011}).

Samples of SMGs are commonly modeled through hydrodynamical simulations (\citealt{Dekel2009}, \citealt{Dave2014}) and semi-analytic models (\citealt{Baugh2005}, \citealt{Swinbank2008}, \citealt{Somerville2012}) within a $\Lambda$CDM cosmology. Since the emission at FIR wavelengths for these sources comes from ultraviolet (UV) light dust-reprocessing, a key process that needs to be addressed by these models is how the dust present in the galaxy absorbs and re-radiates this light. While some of the models relate the parameters controlling the dust attenuation with other galaxy properties and constrain their values with local observations, others make use of radiative transfer codes (\citealt{Silva1998}, \citealt{Jonsson2006} and \citealt{Noll2009}, among others) to compute the FIR luminosity in a self-consistent way. This last method can be very time-consuming when the sample to be modeled comprises thousands or even millions of galaxies, requiring supercomputing facilities.

There is still a controversy between the modeled and observed abundance of SMGs. The lack of enough sources recovered by adopting a Kennicutt \citep{Kennicutt1983} initial mass function (IMF) reported by \citet{Granato2000} was possibly resolved by \citet{Baugh2005}, claiming the need of a flat IMF in starbursts, while still reproducing basic properties of the local galaxy population. This was motivated by the increase both in the total UV light radiated per unit mass of stars and in the yield of metals from core-collapse supernovae, which in turn produces more dust to absorb it. However, the assumption subsequently led to underpredict the $K$-band magnitudes for model SMGs \citep{Swinbank2008}. Like these works, other studies were able to predict the abundance of SMGs via the change in a given assumption or process, but failing in the agreement with other observational statistics like their redshift distribution \citep{Fontanot2007}.

Another possible explanation for the discrepancy between observed and modeled counts is that the low resolution achieved with single-dish telescopes masks multiple sources within a beam. This influence on the number of detected sources can be solved through a follow-up of single-dish detected SMGs with (sub)mm interferometry. \citet{Hayward2013b} and \citet{Cowley2014} have discussed the effects of the beamsize on the counts, using models that give the submm emission of galaxies in a self-consistent way.

The present work is motivated by recent $870\,\mu\textmd{m}$ continuum Atacama Large Millimeter/submillimeter Array (ALMA) observations of the ECDF-S, which was previously surveyed using the Large APEX BOlometer CAmera (LABOCA) on the Atacama Pathfinder EXperiment (APEX) telescope \citep{Siringo2009}. \citet{Karim2013} show that the brightest sources detected in the LABOCA ECDF-S Submillimeter Survey (LESS, \citealt{Weiss2009}) are comprised by emission from multiple fainter sources, namely the ALMA LESS sources (ALESS, \citealt{Hodge2013}). With the aim of exploring the properties of SMGs in this field, and inspired by the now-standard abundance matching technique \citep{Conroy2006}, we perform a ``Count Matching'' approach using lightcones drawn from a semi-analytic model of galaxy formation and evolution. We choose various physical galaxy properties given by the model as proxies for their submm luminosities, assuming a monotonic relationship so that the combined LABOCA plus bright-end ALMA observed number counts are reproduced. After turning the catalogs of galaxy positions and fluxes given by the different proxies into submm maps, we perform a source extraction. With this we study the effects of the observational process on the recovered counts, as well as the galaxy properties derived from the detected sources. For finding the best proxy, we explore the redshift, stellar mass and host halo mass distributions. Once the best proxy is determined, several properties for SMGs (as well as for their descendants) can be predicted.

This paper is organized as follows. Section \ref{sagmodel} gives a brief explanation of the semi-analytic model used. In Section \ref{lightcones} we detail the procedure for constructing lightcones of galaxies, while in Section \ref{countmatching} the count matching technique is described. Section \ref{obsprocess} outlines the observational process for constructing the submm maps of simulated galaxies. In Section \ref{results} results are discussed, ending with a summary in Section \ref{summary}.

\section{Galaxy Formation Model}\label{sagmodel}

We use a combination of a cosmological $N$-body simulation of the concordance $\Lambda$CDM universe and the semi-analytic model of galaxy formation \textsc{SAG}, acronym for Semi-Analytic Galaxies. Details about the semi-analytic model are given in \citet{Cora2006}, \citet{Lagos2008} and \citet{Tecce2010}.

The cosmological simulation was run using the \textsc{GADGET-2} code \citep{Springel2005}. It gives the dark matter halos and their embedded substructures, in a periodic box of $150\,h^{-1}\textmd{Mpc}$ a side, with $640^3$ particles having a mass resolution of $1\times10^9\,h^{-1}\textmd{M}_{\odot}$; 100 snapshots giving information for halos at each epoch are collected, being equally spaced in logarithm of the scale factor between $z=20$ and $z=0$. From those outputs, merger trees are constructed and are then used by \textsc{SAG} to generate the galaxy population. The cosmological parameter values assumed are $\Omega_m=0.28$, $\Omega_{\Lambda}=0.72$, $\Omega_b=0.046$, $H_0=100\,h\,\textmd{km}\,\textmd{s}^{-1}\,\textmd{Mpc}^{-1}$ with $h=0.7$ and $\sigma_8=0.81$. These parameters are consistent with WMAP7 data \citep{Jarosik2011}. Below we present a big picture of the \textsc{SAG} model, referring the reader to the papers cited above for a deeper insight on the physical processes included.

Hot gas in dark matter halos is isothermally distributed, being drawn from the assumed baryon fraction. It cools according to a cooling rate, forming a galaxy disc that follows an exponential distribution. Stars form at a rate that depends on the amount of cold gas that gives origin to them, besides dark matter halo properties (virial radii and velocities). Cold gas inflows \citep{Dekel2009} are not included in the current version of the \textsc{SAG} model.

Stars can form quiescently as well as in starbursts. Starbursts can be triggered by mergers (both major and minor, defined according to the relative masses of the galaxies involved) and disc instabilities, exhausting all the available cold gas instantaneously and contributing to the bulge growth. A merger of two dark matter halos leads to the merging of the galaxies within them. When a merger occurs, satellite galaxies are accreted by a central one after keeping a circular orbit around it, decaying via dynamical friction. Furthermore, a disc becomes unstable when it is massive enough to be dominated by self-gravity, and therefore sensitive to small external perturbations. Both mergers and disc instabilities, besides cold gas accretion, contribute to the growth of a central black hole.

The stellar mass created in each star formation event gives a distinctive number of core-collapse supernovae according to the adopted IMF, which in this model is Salpeter \citep{Salpeter1955}. These supernovae in turn will reheat the cold gas through galactic outflows, transferring it to the hot gas phase. Apart from this kind of feedback, the model considers the heating produced by the active galactic nucleus (AGN) through black hole accretion as responsible of the gas cooling suppression.

Star formation events lead to metal pollution in galaxies, yielded via mass losses through stellar winds in stars having low or intermediate mass, as well as supernovae explosions. The chemical enrichment of the gas affects cooling rates and so star formation.

The degree of efficiency, as well as the fulfillment of several criteria involved in the different physical processes, are regulated by free parameters. These are tuned in order to reproduce a number of $z=0$ observables (see Appendix \ref{sagpred}, where also some of the galaxy properties predicted by the model and not used as constraints are presented). After assuring this, those parameters remain fixed throughout this work, such that none of the values used is dependent on SMG properties (neither from the model itself nor from observational ones).

\section{Constructing Lightcones}\label{lightcones}

\begin{figure}
\begin{center}
\includegraphics[width=0.43\textwidth]{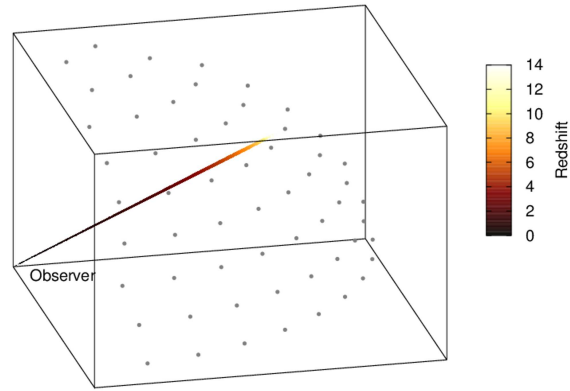}
\caption{Lightcone construction scheme. Consider the first octant of a sphere filled with semi-analytic galaxies and centered on the observer. A given interval in RA and Dec (sky plane) and redshift $z$ (line of sight) is chosen, collecting all the model galaxies that fall within that region (dots coloured by $z$). This section of the sphere is mapped taking different orientations (centers indicated by gray points).}
\label{cone1dir}
\end{center}
\end{figure}

We use the \textsc{SAG} semi-analytic model within the $\Lambda$CDM framework to follow the history of galaxies, obtaining information about them at different epochs which are given by the chosen output redshifts for the model (i.e., the snapshots mentioned in the previous section). We assume that the galaxy sample inside the periodic comoving box of $150\,h^{-1}\,\textmd{Mpc}$ a side is statistically representative of the overall galaxy population in the Universe, such that we can build a simulated universe repeating (if necessary) in our simulated catalogs semi-analytic galaxies at different epochs of their evolution.

For constructing the simulated universe, the observer is placed at the center of a sphere whose radius extends to the highest redshift where a semi-analytic galaxy exists. The whole spherical volume is populated with galaxies taken from the simulated boxes. Since the redshift separation between two consecutive snapshots is smaller than that corresponding to the side of the box, only one box per epoch is needed in the radial direction.

A lightcone consists of all the galaxies that belong to a particular region in the sky and redshift range (see Fig. \ref{cone1dir}), so the choice of a line of sight and limits in right ascension (RA) and declination (Dec) around it are needed. These are chosen as follows\footnote{In the following, we do not include any characteristic of the real sky in the simulated celestial sphere, so none of the selected orientations need to be corrected for the presence of the Milky Way, closeness to bright stars, etc.}. Since our aim is to study the counts in the ECDF-S field, we select the same angular area surveyed by LESS, i.e., $\sim30'\times 30'$. At the redshifts of interest, this area is smaller than the area subtended by the simulation box, so at a given epoch galaxies appear only once in the area covered by the lightcone.

\begin{figure}
\includegraphics[width=0.33\textwidth,angle=270]{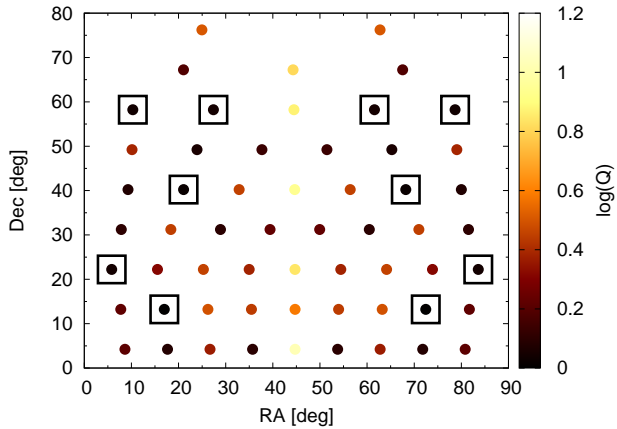}
\hspace{0.2cm}
\caption{Study of the best orientations for the lightcones in RA and Dec, according to the $Q$ quality factor (see text). Orientations indicated in Fig. \ref{cone1dir} are filled with dots representing galaxies, drawn from a cubic grid in the range $z=0$ to 5, taking angular areas of $0.3\,\textmd{deg}^2$. For each orientation, $Q$ is computed (colour code gives its logarithm). The $n$ best orientations are such that $Q$ is lowest (black squares). For this paper $n=10$ lightcones were selected.}
\label{q4grid}
\end{figure}

For a given line of sight, the repetition of a model galaxy at various redshifts is limited as much as possible, since the repetition of simulated structure might lead to a wrong interpretation of how model galaxies are spatially distributed once we turn them into a map. In order to test the amount of sources appearing in the lightcone at more than one epoch, we map the first octant of the sphere in the redshift range $0<z<5$ with 58 lightcones having $0.3\,\textmd{deg}^2$ and orientations with roughly uniform coverage in RA and Dec. Instead of using semi-analytic galaxies, the periodic boxes are filled with a Cartesian grid of dots; they emulate galaxies, so these dots are labeled for keeping control of their frequency of appearance at different epochs. A catalog is then created for each lightcone containing all the ``grid galaxies'' that fall within it, and a quality factor is computed for evaluating the repetition of structures. This factor is defined as
\begin{equation}
Q=\frac{\textmd{\# points appearing more than once}}{\textmd{\# points appearing just once}}, \label{qdef}
\end{equation}
\noindent i.e., the ratio between the amount of sources that appear at more than one epoch and the amount of sources that are not repeated within the lightcone. With this definition, a low repetition of structure is expressed as low $Q$. We choose the $n$ best orientations as those with the lowest $Q$. For this paper, $n=10$ lightcones were selected (see Fig. \ref{q4grid}). The choice of this number of orientations has the aim of testing the effects of cosmic variance over the results. The 58 orientations tested span a range in $Q$ from $1.14$ to $12.66$, with the ten chosen lightcones in the range $Q=1.14-1.24$. Through the analysis of galaxy properties presented in this work, we have checked that there are no trends between $Q$ values and the recovered distributions for these ten lightcones.

This technique for choosing the best orientations for lightcones was tested only up to $z=5$. Since we are interested in galaxy redshift distributions of observed SMGs, which peak at $z\sim2$, this upper limit is adequate for our purposes.

Once a lightcone is constructed, the redshifts of galaxies are computed from their coordinates within it. Identifying the particular snapshots of the simulation to which they belong, several galaxy properties given by \textsc{SAG} can be recovered, e.g., stellar masses, SFRs, etc.

\section{The Count Matching Technique}\label{countmatching}

Motivated by abundance matching techniques (e.g., \citealt{Conroy2006}), we propose a new approach for exploring the properties of SMGs, namely the count matching technique. Here, a given physical galaxy property (or a combination of several ones) is chosen as a proxy for another property whose numerical value is unknown, assuming a monotonic relationship. In our case, the unknowns are the $870\,\mu\textmd{m}$ luminosities of a galaxy sample. We use this monotonic relationship to assign a third property, namely, the $870\,\mu\textmd{m}$ flux density, to the simulated galaxies in such a way that an observational statistics for the latter is reproduced. The chosen observational statistics are the observed galaxy number counts at this wavelength. This allows to explore several other statistics for simulated galaxies selected using the matched fluxes, like redshifts, masses, luminosities, etc. By comparing them with distributions derived from observations, the quality of the proposed proxies can be analyzed.

The selection of properties used in the proxies is motivated by the theoretical understanding of the mechanism that triggers FIR and submm emission in luminous and ultraluminous infrared galaxies (LIRGs and ULIRGs, respectively, \citealt{Sanders1996}), with the SMGs being thought of as the high-$z$ cousins of these sources. These galaxies emit the bulk of their bolometric emission in the FIR, corresponding to absorbed UV light by dust that is thermally re-radiated at FIR wavelengths; this UV light comes mostly from young stars, with a minor contribution from AGN\footnote{\citet{Wang2013} found a fraction of AGN of $17_{-6}^{+16}$ per cent for ALESS SMGs having a rest-frame $0.5-8\,\textmd{keV}$ absorption-corrected luminosity greater than $7.8\times10^{42}\,\textmd{erg/s}$. Moreover, \citet{Laird2010} found that in at least $\sim85$ per cent of GOODS-N SMGs the star formation dominates the FIR emission, via the study of their X-ray spectra.}. Within this picture, it makes sense to propose that the bright submm fluxes measured for these sources are correlated in some way with the stellar mass, for instance, in a very simplistic model motivated by the strong correlation of many galaxy properties with stellar mass and the observation of large stellar masses for SMGs (e.g., \citealt{Michalowski2010}). Or with the dust mass, that partially reprocesses the short wavelength light, or with the star formation rate, as high values give rise to more stars and then more UV emission. Given its low computational cost, the technique offers us the advantage to explore the relationship between several properties (or combinations of them) available from the simulation, and the FIR emission in a galaxy.

We are interested in applying this technique when simulating sources in the ECDF-S, given the well studied SMGs detected using the LABOCA bolometer \citep{Weiss2009} and the recently detected sources using the ALMA interferometer at high submm flux densities \citep{Karim2013}.

\begin{table}
\begin{minipage}{0.5\textwidth}
\begin{center}
\caption{Quantities used as proxies for the rest-frame submm luminosity.}
\begin{tabular}{ccc}
Proxy & Definition\\ \hline \hline
Stellar Mass & $M_{\textmd{stellar}}$\\
SFR & Star formation rate\\
SFR Surface Density & $\textmd{SFR}/r_d^2$ (for $r_d>0$)\\
sSFR & $\textmd{SFR}/M_{\textmd{stellar}}$\\
Dust Mass & $M_{\textmd{dust}}$\\
MDA & $M_{\textmd{dust}}/t_{\textmd{stellar}}$\\
MDD & $M_{\textmd{dust}}\times\textmd{SFR}/r_d^2$ (for $r_d>0$)\\
MDS & $M_{\textmd{dust}}\times\textmd{SFR}$\\
H13 & $M_{\textmd{dust}}^{0.54}\times\textmd{SFR}^{0.43}$\\ \hline
\end{tabular}
\label{proxylist}
\end{center}
\end{minipage}
\end{table}

\begin{figure}
\begin{center}
\includegraphics[width=0.46\textwidth]{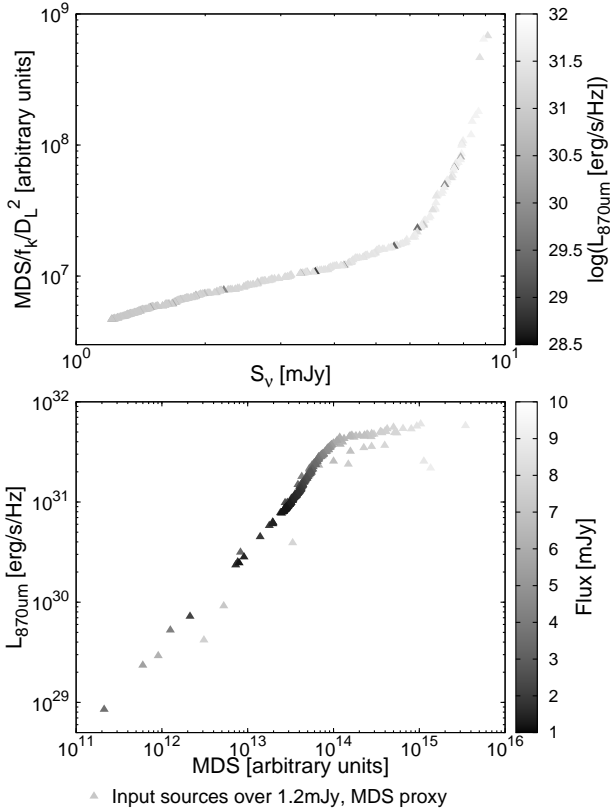}
\caption{The count matching process. Top panel: relation between the $870\,\mu\textmd{m}$ flux and the flux-analog quantity $y=\textmd{Proxy}/f_k/D_L^2$ for the MDS proxy. Bottom panel: recovered relation between rest-frame $870\,\mu\textmd{m}$ luminosity and proxy value. For clarity, this is shown only for sources brighter than $1.2\,\textmd{mJy}$ for a given lightcone.}
\label{proxytot}
\end{center}
\end{figure}

The steps for this approach are the following:

\begin{enumerate}
\item Select a galaxy sample. We take all the galaxies within each lightcone with stellar mass over $10^8\,\textmd{M}_{\odot}$. This lower limit is imposed by the mass resolution of the underlying cosmological simulation; the luminosity functions obtained for our semi-analytic galaxies can be considered complete till that limit (see \citealt{Ruiz2014}).

\item Choose a physical galaxy property as a proxy for the $870\,\mu\textmd{m}$ rest-frame luminosity, assuming that sources having higher values of the property have higher luminosities. With this we are imposing a monotonic relationship between proxy and luminosity, without placing any restriction about the exact shape of this relation (which can vary depending on the chosen proxy). We test several properties given by \textsc{SAG}, as well as combinations of two or more. The chosen properties are: stellar mass ($M_{\textmd{stellar}}$), star formation rate (SFR), cold gas mass ($M_{\textmd{cold gas}}$), disc scale length ($r_d$), dust mass ($M_{\textmd{dust}}$), stellar age ($t_{\textmd{stellar}}$) and cold gas phase metallicity ($\log(\textmd{O/H})+12$). The proxies constructed with these properties are defined in Table \ref{proxylist}. The H13 proxy corresponds to the eq. 15 of \citet{Hayward2013}, i.e. $M_{\textmd{dust}}^{0.54}\times\textmd{SFR}^{0.43}$, being the best fit over their simulated galaxies. For the MDA, MDD, MDS and H13 proxies, the dust mass is computed assuming $M_{\textmd{dust}}\propto M_{\textmd{cold gas}}\times(\log(\textmd{O/H})+12)$. This is inspired by the assumption of a dust-to-gas mass ratio proportional to the metallicity,adopted by the radiative transfer code \textsc{GRASIL} \citep{Silva1998}.

\item For a given property or a combination of properties, let the proxy value, adjusted for cosmological distance, be known as
\begin{equation}
y=\frac{\textmd{Proxy}}{f_k\,D_L^2}, \label{yfactor}
\end{equation}
\noindent and sort the values for all galaxies in increasing order. Here $D_L$ is the luminosity distance of each galaxy and $f_k$ is a factor giving the $k$-correction corresponding to its redshift \citep{Hogg2002} at rest-frame $870\,\mu\textmd{m}$, defined in terms of redshift and monochromatic luminosities (in the $L_{\nu}$ formalism) as
\begin{equation}
f_k=\frac{L_{870\,\mu\textmd{m}}}{(1+z)\,L_{870\,\mu\textmd{m}/(1+z)}}.
\end{equation}
\noindent Note that the definition of $y$ introduced above turns it into an analog to a flux density (despite its arbitrary units, which vary according to the proxy), since the rest-frame $870\,\mu\textmd{m}$ luminosity can be related to the observed $870\,\mu\textmd{m}$ flux density through
\begin{equation}
S_{870\,\mu\textmd{m}}=\frac{L_{870\,\mu\textmd{m}}}{4\pi\,f_k\,D_L^2}.
\end{equation}
\noindent The $k$-correction is recovered using a template for Arp220 \citep{Blain1999}, which is a typical ULIRG. Additionally, we explore the effect of assuming that the negative $k$-correction eliminates completely the diminution of the submm flux due to distance; in this approach, the amount of submm flux assigned to each galaxy only depends on the value of the property or combination of properties selected as proxy.
\item Assign a submm flux to each galaxy according to its value of $y$, such that sources with higher values will have higher fluxes. This is the key step in the count matching process. Submm fluxes are drawn from a Monte Carlo simulation following the observed cumulative number counts, where we have combined the LABOCA counts at low fluxes \citep{Weiss2009} and ALMA counts \citep{Karim2013} over $8\,\textmd{mJy}$; the choice of ALMA counts only for the bright end is because we want to test whether we are able to recover the LABOCA counts after simulating the observational process (including blending), while avoiding biases in the ALMA counts arising from targeting only bright LESS sources and not other regions in the ECDF-S with signal-to-noise ratios slightly lower than the LESS threshold \citep{Karim2013} (we tested the effect of switching counts between ALMA and LABOCA data at different fluxes in the range $7.5-8.5\,\textmd{mJy}$, which does not lead to significantly different results). We translate this combination of cumulative counts into differential counts, and use one realization of the Monte Carlo simulation as the random fluxes that follow these counts.

The assignment of submm fluxes for galaxies in the lightcone can be summarized as
\begin{equation}
\int_{S_{\nu}'}^{\infty}\!n(S_{\nu}){dS_{\nu}}=\int_{Y}^{\infty}\!n(y){dy} \label{cm}
\end{equation}
\noindent where $S_{\nu}$ is the flux density at $870\,\mu\textmd{m}$ and $n(S_{\nu})$ is the amount of galaxies having flux densities between $S_{\nu}$ and $S_{\nu}+dS_{\nu}$ (and similarly for $n(y)$). $Y$ and $S_{\nu}'$ stand for particular values of $y$ and $S_{\nu}$. Then $Y(S_{\nu}')$ gives the transformation from flux to proxy, which in turn can be used to recover the numerical value of the submm luminosity for each galaxy.
\end{enumerate}

To illustrate the procedure, in Fig. \ref{proxytot} top panel we show the relation between the $870\,\mu\textmd{m}$ flux, which comes from the Monte Carlo simulation, and the flux-analog quantity $y$ for the MDS proxy, which comes from the model (for clarity, this is only shown for sources over $1.2\,\textmd{mJy}$ for a given lightcone). Once the $870\,\mu\textmd{m}$ flux is assigned to a given galaxy, the SED template can be used to recover the numerical value of its rest-frame $870\,\mu\textmd{m}$ luminosity. The shape of the monotonic relation between this luminosity and the proxy value for MDS is shown in the bottom panel; the trend is robust, despite some outliers having combinations of extreme proxy and redshift values.

Proxies proposed in this work then have simple, one-to-one dependences with rest-frame $870\,\mu\textmd{m}$ luminosity, for a stellar mass limited sample of model galaxies. No additional selections are included at this stage, except for sources where a proxy has an undefined value (e.g., removing galaxies without disc in SFR surface density and MDD proxies). Definitely, these proxies can be refined selecting only a subset of the stellar mass limited sample, for instance removing passive galaxies because of their low SFR; we leave these improvements for a future work.

The advantage of using this simple recipe instead of a full radiative transfer is the considerably lower computational cost, when applied to big samples of galaxies. However, it also has limitations. In the abundance matching formalism, the halo mass function must be correctly predicted by the $N$-body simulation, as well as the stellar mass function should be well derived from the observations. In our approach, we are assuming that the bright-end ALMA and faint-end LABOCA counts are correct, an observable that does not provide redshift information. Because submm luminosities are difficult to model directly, we choose a proxy for the rest-frame submm luminosity for each model galaxy, and use its redshift and an appropriate SED template to predict its submm flux analog. We then perform an abundance-type matching in the space of observed vs predicted submm fluxes following Eq. \ref{cm}. This determines the redshift distribution of the brightest sources (i.e. SMGs) as a function of flux, which in turn influences their clustering and therefore the amount of blending that occurs before bright-end LABOCA counts are measured.

\section{Simulating the Observational Process}\label{obsprocess}

\begin{figure}
\begin{center}
\includegraphics[width=0.4\textwidth]{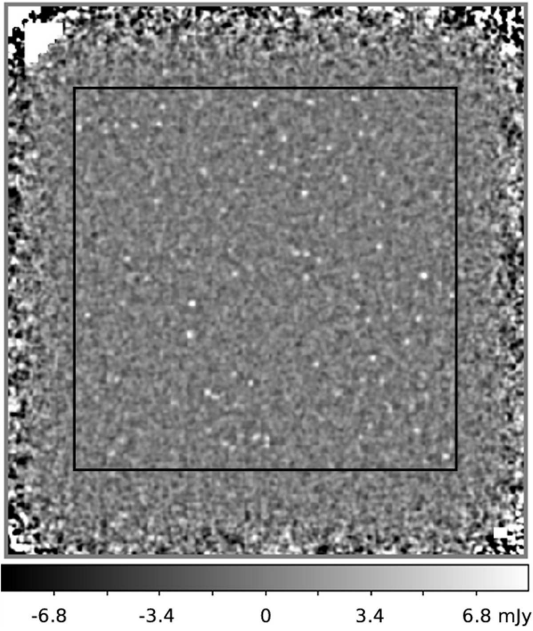}
\caption{Example of a simulated map, corresponding to a given lightcone in the MDS proxy. The black square encloses the region where simulated sources are injected ($\sim30'\times 30'$). Grayscale gives submm flux per pixel. The map resolution is $6''/\textmd{pix}$.}
\label{mapmds6}
\end{center}
\end{figure}

We turn the catalogs of galaxy positions and fluxes given by the different proxies into submm maps that include a modeling of the observational process.

First, simulated sources are injected in a noise map having a spatial resolution of $19.5''$ (as the LABOCA beamwidth). The resulting map is beamsmoothed using a $90''$ Gaussian kernel, for removing low spatial frequency structures as is commonly done for observational data (e.g., \citealt{Weiss2009}). The latter map is then substracted from the former, and the result is convolved with a $19.5''$ Gaussian kernel, giving rise to a map resolution of $27.6''$. The beamsmoothing process produces a decrease in the source fluxes by $\sim8.6$ per cent, which are hence rescaled.

One map is constructed for each orientation and proxy. Injected sources span a region of $\sim30'\times 30'$. They are taken from the simulated lightcones, being the $N$ brightest sources of each catalog. We choose $N=5\times10^3$, as it gives source fluxes well below the values reported by observations, while the computing time is reasonable (we checked that a change in one order of magnitude in $N$ does not have a significant effect on the statistics recovered from the maps). An example of these simulated maps is given in Fig. \ref{mapmds6}, indicating the region where model sources are injected as a black square.

We perform a source extraction as done for maps obtained through observations. With this we study the effects of the observational process on the recovered counts, as well as the galaxy properties derived from the counterparts of detected sources for each proxy.

\begin{figure}
\begin{center}
\includegraphics[width=0.46\textwidth]{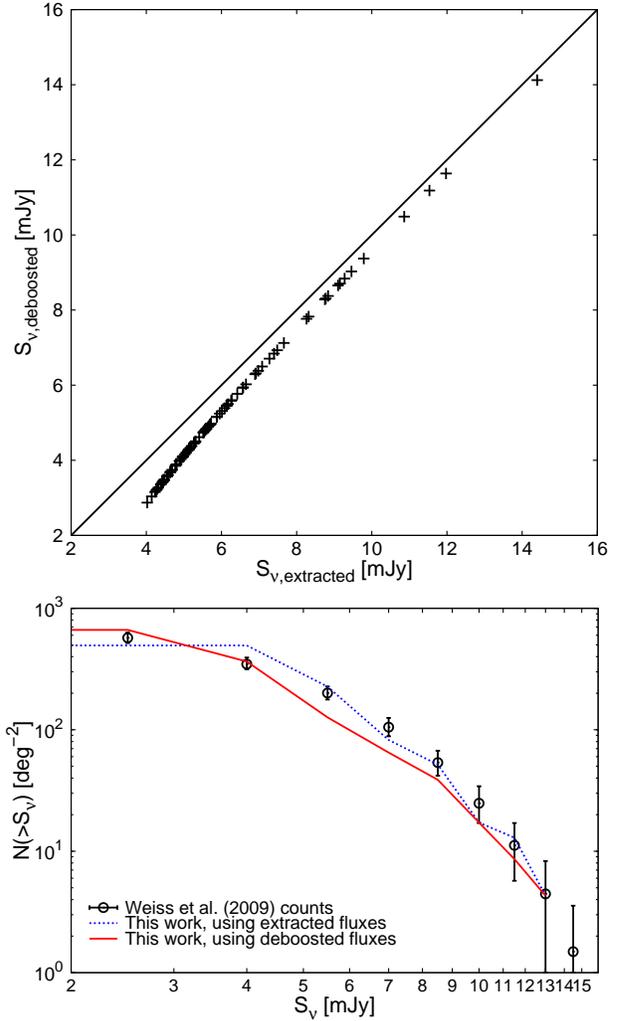}
\caption{Source extraction and flux deboosting methods used in this work, applied to the actual LESS map \citep{Weiss2009}. Top panel: comparison between extracted and deboosted fluxes for our detected sources. Bottom panel: cumulative number counts recovered in each case, compared to actual data for this field \citep{Weiss2009}.}
\label{countsless}
\end{center}
\end{figure}

Sources are extracted using \textsc{SExtractor} \citep{Bertin1996}, for a limit in S/N of 3.8. This cut is chosen following \citet{Weiss2009}, according to the expected number of false detections. The source extraction comprises two iterations, where in the second one the sources detected in the first iteration are removed from the map and the noise level is recalculated. We consider even sources extracted outside the limits given by the injected sources.

As our resulting source catalog is limited by S/N, we need to correct the extracted fluxes for boosting effects: the systematic enhancement on the measured fluxes coming from noise introduces a bias in the number of sources exceeding the chosen limit \citep{Scott2008}. For simulated maps, the deboosting correction is found comparing the extracted flux of a given source with the flux of its brightest counterpart in the input catalog (see Section \ref{counts}) and then using the fitting function $S_{\nu,\textmd{deboosted}}=(S_{\nu,\textmd{extracted}}+A)^{1/2}$, with $A$ the free parameter. This function is chosen such that it has a cutoff at fluxes near the S/N limit, and tends to the equality at high fluxes. Taking these deboosted fluxes, cumulative submm number counts are obtained for each simulated map, including a proper correction for completeness taken from \citet{Weiss2009}.

In order to validate these methods of source extraction and flux deboosting, we apply them to the actual LESS field \citep{Weiss2009}: first, we extract sources from the actual LESS map down to $\textmd{S/N}=3.8$ with \textsc{SExtractor}, and secondly we apply the flux deboosting correction to the extracted sources. Since in this case the map comes from real observations, we do not have an input catalog for it, so the $A$ parameter is found computing the median value among all the simulated maps. The resulting deboosted fluxes and counts are shown in Fig. \ref{countsless}. Compared to \citet{Weiss2009} data, we recover a lower amount of sources in the flux range $\sim4-8\,\textmd{mJy}$ by a factor $\sim1.5$. However, this may come from their different deboosting flux technique, which makes use of a $P(D)$ analysis.

We have tried different parameters for \textsc{SExtractor}. In a first extraction with conservative parameters (\textsc{DETECT\_MINAREA}=5 plus defaults) and only one iteration, we obtained $\sim20$ per cent less sources than reported by \citet{Weiss2009}. Trying with a grid of parameters in \textsc{DETECT\_MINAREA}, and also modifying the parameters related to the deblending of sources (\textsc{DEBLEND\_NTHRESH} and \textsc{DEBLEND\_MINCONT}), we increased the matching with the original catalog. Finally, after extracting the sources from the first iteration, we re-run \textsc{SExtractor} on the residual map. This procedure assures us a good agreement with the catalog from \citet{Weiss2009}, thus minimizing differences in the final counts arising from different detection methods.

\section{Results and Discussion}\label{results}

\begin{figure*}
\begin{center}
\includegraphics[width=0.66\textwidth]{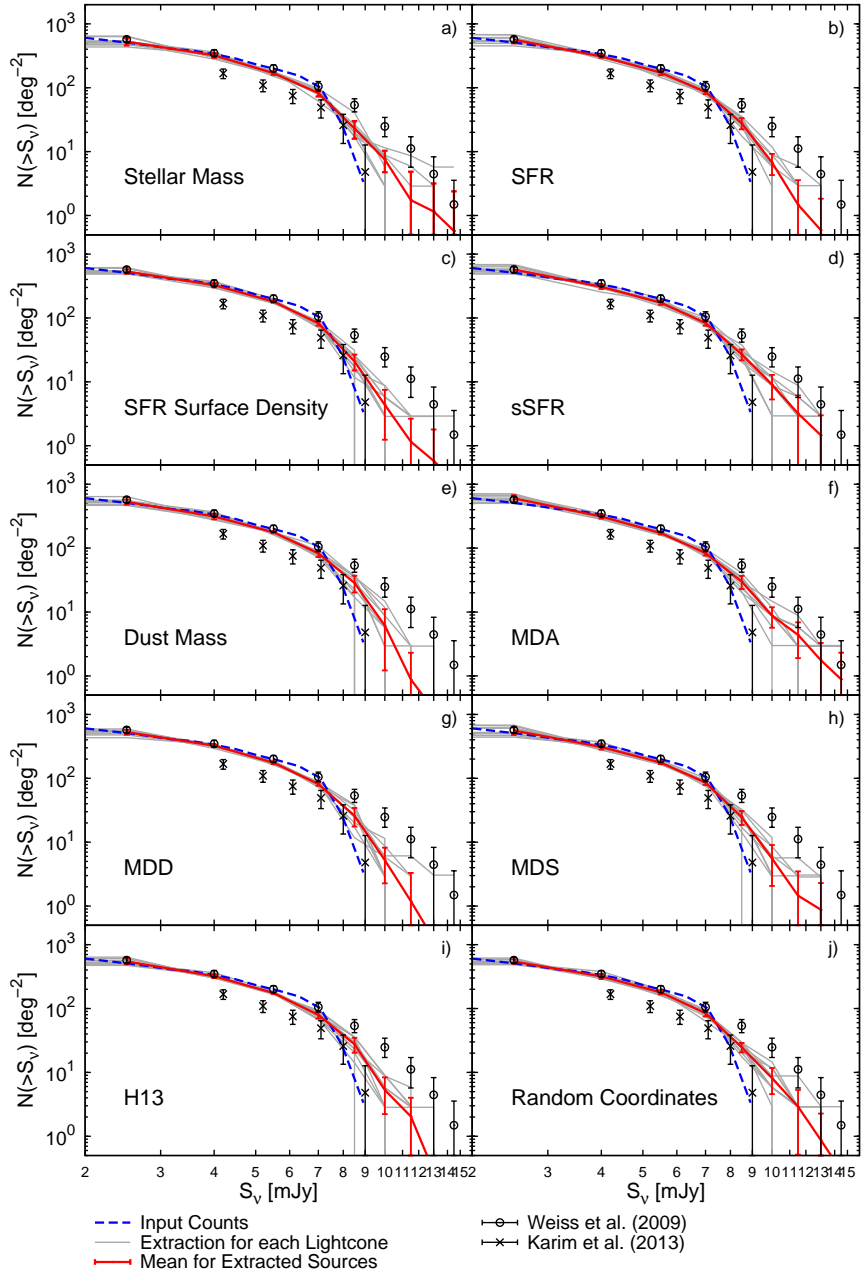}
\caption{Cumulative number counts at $870\,\mu\textmd{m}$ obtained for the different proxies: a) stellar mass, b) SFR, c) SFR surface density, d) specific SFR (sSFR), e) dust mass, f) dust mass divided by stellar age (MDA), g) dust mass times SFR divided by the square of the disc scale length (MDD), h) dust mass times SFR (MDS), and i) eq. 15 of \citet{Hayward2013} (H13). Blue thick dashed lines give the input counts for each proxy, using all the input galaxy fluxes given by the count matching. Gray thin solid lines correspond to the counts recovered for the ten lightcones after passing through the observational process, i.e., obtained from sources extracted from the simulated maps. Red thick solid lines connect mean values for simulated sources at each flux level, with standard deviations as errorbars. As a comparison, observational data are shown from LABOCA \citep{Weiss2009} and ALMA studies \citep{Karim2013}. Additionally, counts obtained for input sources with randomized positions in the simulated map are given in panel j). In all panels, there are lines of sight giving counts consistent with those derived from LABOCA observations.}
\label{countscones}
\end{center}
\end{figure*}

We are interested in those proxies where the simulated sources that follow the observed LABOCA plus bright-end ALMA counts, after going through the observational process, give 1) counts comparable to LABOCA data, 2) redshift distributions consistent with observational values, 3) other properties in agreement with observations (including clustering, stellar mass, host halo mass, etc). For the last two requirements, we compare model distributions with different surveys from the literature.

Note that it is possible that the properties are affected by the environment of ECDF-S, as this field appears to be underdense when compared to other deep fields probed with submm wavebands (see \citealt{Weiss2009} discussion, but also \citealt{Chen2013}, who find no difference with other fields but for a smaller area). However, we find that there are only minor changes in the proxy vs luminosity at $870\,\mu\textmd{m}$ relation (see Fig. \ref{proxytot}) between lightcones of different density. In particular, this relation changes only by $\sim10$ per cent between the different lightcones for the MDS proxy.

\subsection{Number Counts}\label{counts}

\begin{table*}
\begin{minipage}{\textwidth}
\begin{center}
\caption{Properties from the counts derived for each proxy, using sources extracted from the simulated maps (extraction down to the $3.8\sigma$ detection level). Fractions are mean values over the ten lightcones, reporting also standard deviations.}
\begin{tabular}{cccc}
\multirow{2}{*}{Proxy} & \multirow{2}{*}{$\chi^2/\textmd{dof}$ \footnote{Reduced $\chi^2$ for each proxy, with $\chi^2$ calculated according to Eq. \ref{chi2} and $\textmd{dof}$ the degrees of freedom (i.e. number of flux bins).}} & Mean fraction of & Mean fraction of\\
& & non-matches \footnote{Mean fraction of sources without having a counterpart in the input catalog (over all the extracted sources).} & multiples \footnote{Mean fraction of sources having more than one counterpart in the input catalog (over the extracted sources that have at least one counterpart).}\\ \hline \hline
Stellar Mass & 3.598 & $0.223\pm0.027$ & $0.113\pm0.044$\\
SFR & 3.120 & $0.250\pm0.032 $ & $0.105\pm0.028$\\
SFR Surface Density & 3.689 & $0.232\pm0.029$ & $0.090\pm0.033$\\
sSFR & 3.617 & $0.237\pm0.026$ & $0.123\pm0.028$\\
Dust Mass & 3.536 & $0.243\pm0.028$ & $0.091\pm0.021$\\
MDA & 3.400 & $0.228\pm0.021$ & $0.141\pm0.034$\\
MDD & 3.642 & $0.232\pm0.029$ & $0.102\pm0.033$\\
MDS & 3.558 & $0.244\pm0.025$ & $0.096\pm0.028$\\
H13 & 3.599 & $0.245\pm0.041$ & $0.089\pm0.034$\\
Random Coordinates & 3.470 & $0.248\pm0.021$ & $ 0.093\pm0.027$\\ \hline
\end{tabular}
\label{10cones}
\end{center}
\end{minipage}
\end{table*}

Fig. \ref{countscones} shows the cumulative submm number counts obtained for the different proxies, compared to the input number counts (i.e., before passing through the observational process) and to those extracted from LABOCA and ALMA observations. Compared to the observed counts, sources extracted from the simulated maps give counts closer to LABOCA data.

In order to quantify the goodness of each proxy at this step, we compare their $\chi^2$ values given by
\begin{equation}
\chi^2=\sum_{S_{\nu}}\frac{[(dN/dS)_{\textmd{LABOCA}}-\overline{(dN/dS)}]^2}{[\delta (dN/dS)_{\textmd{LABOCA}}]^2+[\delta\overline{(dN/dS)}]^2}, \label{chi2}
\end{equation}
\noindent where $(dN/dS)_{\textmd{LABOCA}}$ corresponds to the observed LABOCA differential counts at a given flux density, $\overline{(dN/dS)}$ gives our mean differential counts over the ten lightcones, and $\delta (dN/dS)$, $\delta\overline{(dN/dS)}$ values are the reported uncertainties in LABOCA counts and scatter in our simulated counts, respectively. Since sources injected in the maps follow the combined bright-ALMA plus faint-LABOCA counts by construction, this $\chi^2$ definition allow us to quantify the influence coming from a) the observational process (blending, noise, etc) and b) clustering given by the model (as different proxies give different spatial distribution for bright sources).

The reduced $\chi^2$ is presented for all proxies in Table \ref{10cones}, second column. There, as well as in Fig. \ref{countscones}, we also show the results for the case where the spatial distribution of input sources along each of the ten maps is random, i.e., does not come from any proxy and so no clustering is provided (note that the number of sources injected in each map is the same as used for the proxies, as well as their submm flux distribution). All proxies are quite similar, so this criterion is not enough to choose a proxy as the best. A further analysis of the predicted distribution of other galaxy properties may discriminate better between proxies. Even the counts obtained for input sources with randomized positions in the simulated map are consistent with those where the count matching process was applied (which include realistic galaxy clustering provided by the semi-analytic model). This finding indicates that the clustering has a minor influence when determining the cumulative submm number counts.

For recovering the properties of model galaxies, we perform a cross-match between all the extracted sources and the input galaxies having submm fluxes down to $1.2\,\textmd{mJy}$ (hereafter the input catalog). A search radius of $13.8''$ is used, corresponding to the beam radius after the map processing (see Section \ref{obsprocess}). When more than one input galaxy falls within the search radius around a given extracted source, we refer to it as a multiple source composed by blended galaxies; otherwise, we call it a single source.

\begin{figure}
\begin{center}
\includegraphics[width=0.44\textwidth]{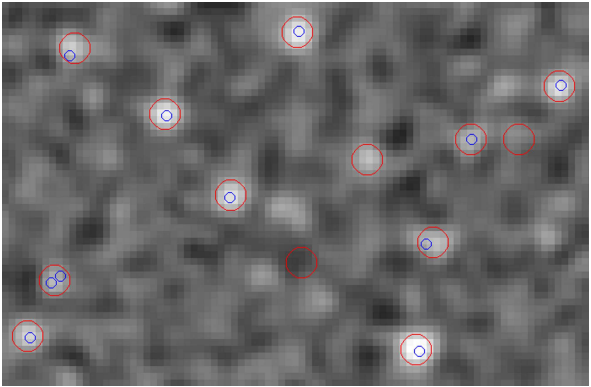}
\caption{Small section of the simulated map shown in Fig. \ref{mapmds6}, illustrating the cross-match between extracted and input sources. Sources extracted from the map lie in the centers of the red big circles, which have a radius of $13.8''$; this is the search radius used for finding the counterparts of these sources in the input catalog down to $1.2\,\textmd{mJy}$, which are shown as blue small circles. They allow us to explore the predicted properties for extracted sources in each lightcone and proxy.}
\label{mapmds6_allc}
\end{center}
\end{figure}

Fig. \ref{mapmds6_allc} illustrates this with a small region within the map shown in Fig. \ref{mapmds6}. Note that among the extracted sources, besides singles and multiples there can be also sources that do not have a counterpart in the input catalog. Although most of these non-matched sources have S/N ratios around the chosen limit for extraction and can be considered as spurious sources whose flux is boosted by the map noise, some of them have S/N over 5.

A fraction of $\sim22-25$ per cent of the extracted sources do not have counterparts (see Table \ref{10cones}, third column); in the following, we remove these sources from our analyses. The fraction of multiple sources across the proxies, in the flux range given by the recovered counts, goes from $\sim9$ to $\sim14$ per cent (see Table \ref{10cones}, fourth column). We have checked that the coordinates of counterparts do not follow a preferential direction in the map when compared to the coordinates of extracted sources.

\subsection{Recovered Redshift Distributions for Detected Sources and $k$-correction Effects}\label{zdistrib}

As shown above, all the proposed proxies (even taking random positions for galaxies in the sky) can successfully reproduce the observed LABOCA counts. We then explore other predicted properties of bright model galaxies, to find out which (if any) proxies for submm luminosity are plausible.

The properties of extracted sources are recovered assigning to them the galaxy properties of their counterpart(s) in the input catalog. Including in our analysis the properties of all the input sources within the search radius down to $1.2\,\textmd{mJy}$ allows us to compare the distribution of recovered properties for each proxy with statistics derived from observations in the literature, even from follow-ups of single-dish detected SMGs with interferometry. If we restrict the comparison to single-dish detected SMGs, and so assign to a extracted multiple source the properties of the brightest model galaxy within the search radius, there are no significant differences in the recovered distributions compared to the inclusion of all counterparts. This shows that the observational process of source extraction does not impose systematic biases in this respect. In addition, we can compare the observations in the literature with the distributions obtained for input sources brighter than $3.8\,\textmd{mJy}$, i.e., the properties of the brightest model galaxies in submm before going through the observational process.

\begin{figure*}
\begin{center}
\includegraphics[width=0.55\textwidth]{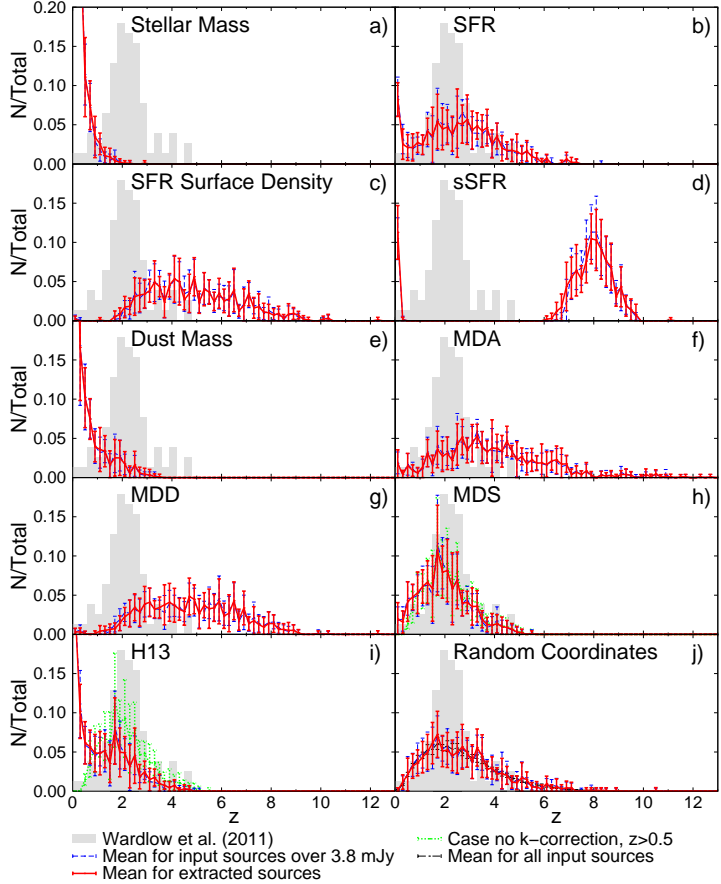}
\caption{Mean redshift distributions over the ten lightcones obtained for the different proxies, taking all the sources extracted from simulated maps having counterparts in the catalog of input sources down to $1.2\,\textmd{mJy}$, considering all the input sources that lie within a search radius of $13.8''$ (red thick solid lines). This is compared to the mean histogram obtained taking all the input sources brighter than $3.8\,\textmd{mJy}$ (blue thin dashed lines). For each proxy the distribution of photometric redshifts obtained by \citet{Wardlow2011} for 78 LESS sources is overplotted (gray histograms). The submm flux for each galaxy in the input catalogs, which give origin to the simulated maps, is obtained according to Eq. \ref{cm} (see also Eq. \ref{yfactor}), where the $k$-correction factor is given by an Arp220 template spectrum. The MDS proxy is the only one that is in good qualitative agreement with the observed distribution for the ECDF-S. The H13 proxy only gives a good agreement when model sources having $z<0.5$ are excluded in the count matching process (green thick dotted lines); in this case, $k$-correcting the spectra does not introduce a significant change in the distribution, and MDS remains as a good proxy. Additionally, the last panel shows the distribution recovered when a random assignment of submm luminosities is done for model sources, injected in a map following a random spatial distribution. In this panel we also include the distribution for the whole galaxy population in the simulated lightcones (black thin dot-dashed line).}
\label{zdistribinp}
\end{center}
\end{figure*}

Fig. \ref{zdistribinp} shows the mean redshift distributions for the extracted sources in each proxy. For all proxies, there is no significant difference between them and the distributions for the input brightest sources. The proxy that gives the closest match with the observed distribution (i.e., \citealt{Wardlow2011} for the ECDF-S) is MDS, giving mean values around $z=2$. Although the SFR proxy gives a distribution with similar shape to the observational one, it has the problem of including a significant fraction of very low-$z$ sources. Hence, even though the MDS proxy prediction is much broader compared to the observed distribution, it provides an acceptable agreement for such a simple model where the rest-frame $870\,\mu\textmd{m}$ luminosity is assumed to be related to galaxy properties in a direct dependence (see Fig. \ref{proxytot}), suggesting that both dust mass and SFR might play an important role in the process of FIR emission. This dependence can surely be improved.

For the brightest sources in the submm, however, redshift distributions also depend on the $k$-correction factor assumed for the model galaxies, as it affects the assignment of submm fluxes to each galaxy in the sample and thus the flux of sources detected in the simulated maps; similarly, the distributions may change if very low-$z$ sources are excluded in the count matching process. Assuming that the negative $k$-correction eliminates completely the dilution due to distance, the amount of submm flux assigned to each galaxy only depends on the value of the property selected as proxy; we call this case ``no $k$-correction'', and explore it particularly for the MDS and H13 proxies. The predicted distributions are shown as green histograms in panels h) and i) of Fig. \ref{zdistribinp} respectively, for comparison with the fiducial case. As it is shown, H13 becomes a good proxy only if model sources having $z<0.5$ are excluded in the count matching process. In this case, there are no significant modifications to the shape of the distribution when applying the $k$-correction to model spectra.

\begin{figure*}
\begin{center}
\includegraphics[width=1.\textwidth]{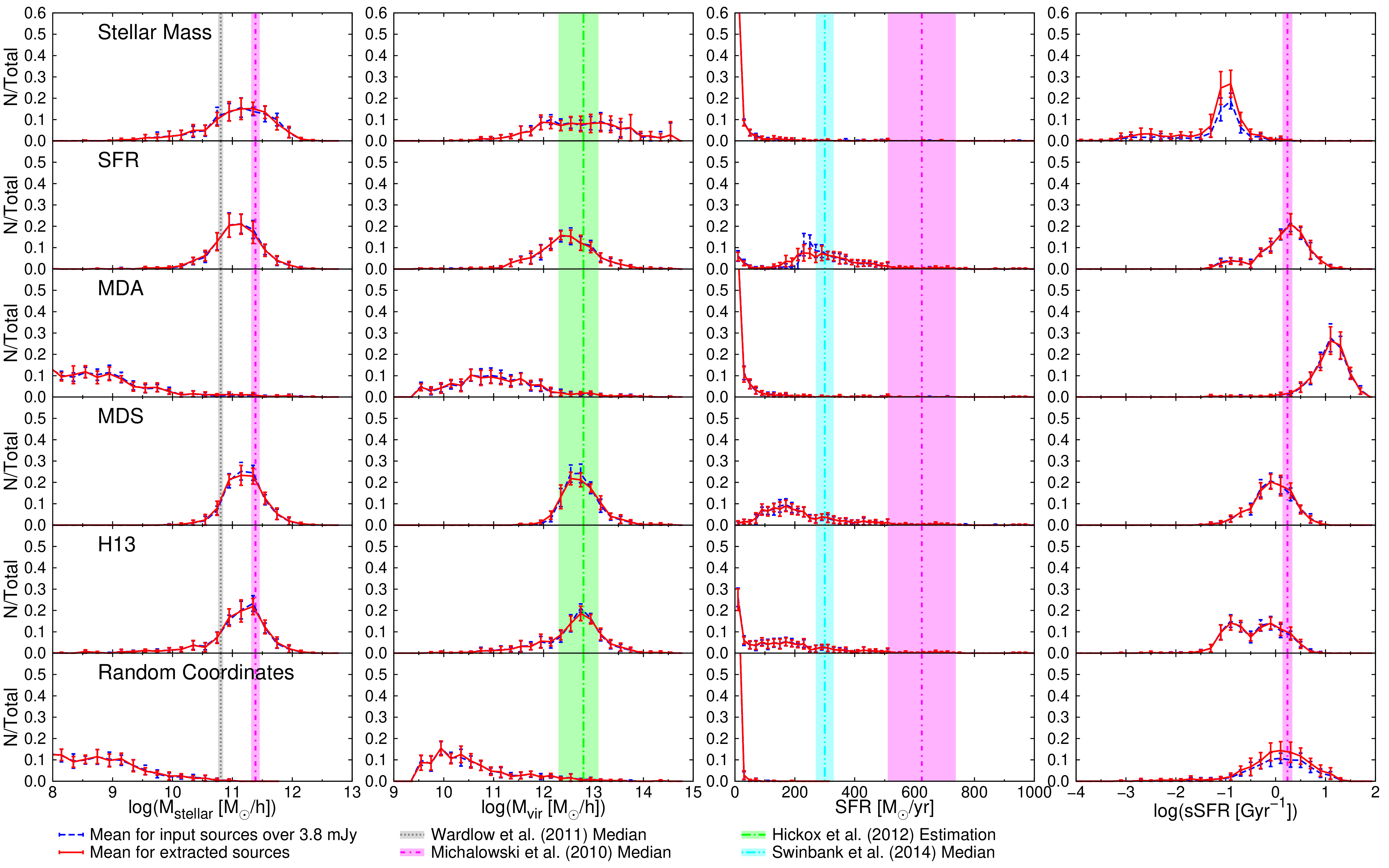}
\caption{Mean distributions over the ten lightcones of several properties obtained for five selected proxies: stellar mass, SFR, dust mass divided by stellar age (MDA), dust mass times SFR (MDS), and eq. 15 of \citet{Hayward2013} (H13). Sources extracted from simulated maps having counterparts in the catalog of input sources down to $1.2\,\textmd{mJy}$ are taken, considering all the input sources that lie within a search radius of $13.8''$ (red solid lines). Properties are compared to the mean histograms obtained taking all the input sources brighter than $3.8\,\textmd{mJy}$ (blue dashed lines). Standard deviations for model distributions are shown as errorbars. In columns from left to right, the predicted properties are: stellar mass, compared to the median values obtained by \citet{Wardlow2011} and \citet{Michalowski2010}, in vertical gray dotted and magenta dot-dashed lines respectively; virial halo mass, compared to the \citet{Hickox2012} estimation, as a vertical green long-dashed-dotted line; SFR, compared to the \citet{Michalowski2010} median value as vertical magenta dot-dashed line and to the \citet{Swinbank2014} median value as a vertical cyan dot-dot-dashed line; and specific SFR, compared to \citet{Michalowski2010} median value, again in a vertical magenta dot-dashed line. Standard errors for observational values are presented as coloured shaded regions. Bottom panels show the distributions recovered when a random assignment of submm luminosities is done for model sources, injected in a map following a random spatial distribution.}
\label{distribs1}
\end{center}
\end{figure*}

The dependence of the submm flux density with dust mass and SFR was addressed in detail by \citet{Hayward2011}, using the results of hydrodynamical simulations of isolated disc and merging galaxies connected to a dust radiative transfer code. \citet{Hayward2011} results motivated the fitting function used by \citet{Hayward2013} in the assignment of submm fluxes. They found that this relation is accurate to within $0.3\,\textmd{dex}$ in the redshift range $\sim1-6$, but severely underpredicts the galaxy fluxes for $z\lesssim0.5$. Our results are in line with those findings, as the H13 proxy reproduces the observed redshift distribution only when discarding low-redshift sources during the flux assignment. Over $z>0.5$, this proxy works applying or not the $k$-correction factor. It is remarkable that the MDS proxy gives a good agreement with observations even when sources at $z<0.5$ are excluded in the count matching process, and independent of the assumption for the shape of the galaxy spectrum.

Besides the proposed proxies, we explore the redshift distribution obtained when a random assignment of submm luminosities is done for model sources, which are injected in the simulated map with a random spatial distribution. As is shown in panel j) of Fig. \ref{zdistribinp}, the recovered distribution is wider than the observed for SMGs and reaches its maximum around $z=2$, resembling the statistics for the whole galaxy population in the simulated lightcones.

We apply some robustness tests to our technique. First, we test the effects of adding scatter to the proxy, in light of the findings by \citet{Moster2010} and \citet{Behroozi2010} about how scatter can bias the results of the abundance matching. We explore this using a Gaussian distribution centered in the value given by the model and taking a standard deviation of 30 per cent (this choice is close to typical errors in galaxy properties derived from observations, as stellar mass or SFR). After this, the trend between proxy and rest-frame $870\,\mu\textmd{m}$ luminosity is still robust. The recovered redshift distributions are consistent with the ones obtained without considering the scatter, with their medians varying only 0.15 units at most (except for the sSFR proxy, where the median redshift decreases by 0.9 units).

In addition, we test the effects that come from using models that give poorer fits to the observed galaxy population. We achieve this modifying two key features in \textsc{SAG}: a) excluding AGNs in the whole model, and b) using the \citet{DeLucia2004} prescription for star formation (fiducial is \citealt{Croton2006}, which unlike the former involves a cold gas mass density threshold below which there is no star formation). Both tests were done keeping the same model parameters as in the fiducial case.

Excluding AGNs changes dramatically the SMG redshift distribution for the MDS proxy, which now peaks at $z=0.3$. For the stellar mass, dust mass and H13 proxies it gives no sources over $z=1$, while for the SFR proxy it gives $\sim3$ times more low-$z$ sources in proportion. For this model, none of our proposed proxies gives an acceptable redshift distribution when compared to observational data.

The model with the alternative prescription for star formation introduces changes in the SMG redshift distributions for some of the proxies. For the MDS and H13 proxies it gives 2 and 1.6 times more low-$z$ sources in proportion, respectively. For the dust mass proxy it gives 1.5 times more $z=0$ sources and essentially zero sources over $z=1$. For the SFR surface density proxy it gives no sources below $z=3$, while for the MDD proxy there are no SMGs below $z=4$.

Finally, we tested the effects on the recovered distributions of using the median SED for observed SMGs found by \citet{Michalowski2010} to compute $k$-corrections, finding no significant differences with respect to the fiducial model except for the sSFR proxy, where the median redshift decreases by 2 units. For the H13 proxy, an extra peak emerges in the predicted redshift distribution at $z=0.8$. We also tried \citet{Magdis2012} SED templates as they allowed \citet{Bethermin2012} to reproduce the counts from the mid-IR to the mm domain using an empirical model. Using these templates we find no significant differences with respect to the fiducial model. For instance, in the SFR, MDS and H13 proxies the median redshift moves to a lower value by $\sim0.5$ units. To sum up, the changes from modifications to the model produce small changes in the results as long as the $z=0$ model galaxies are consistent with observations.

\begin{figure*}
\begin{center}
\includegraphics[width=1.\textwidth]{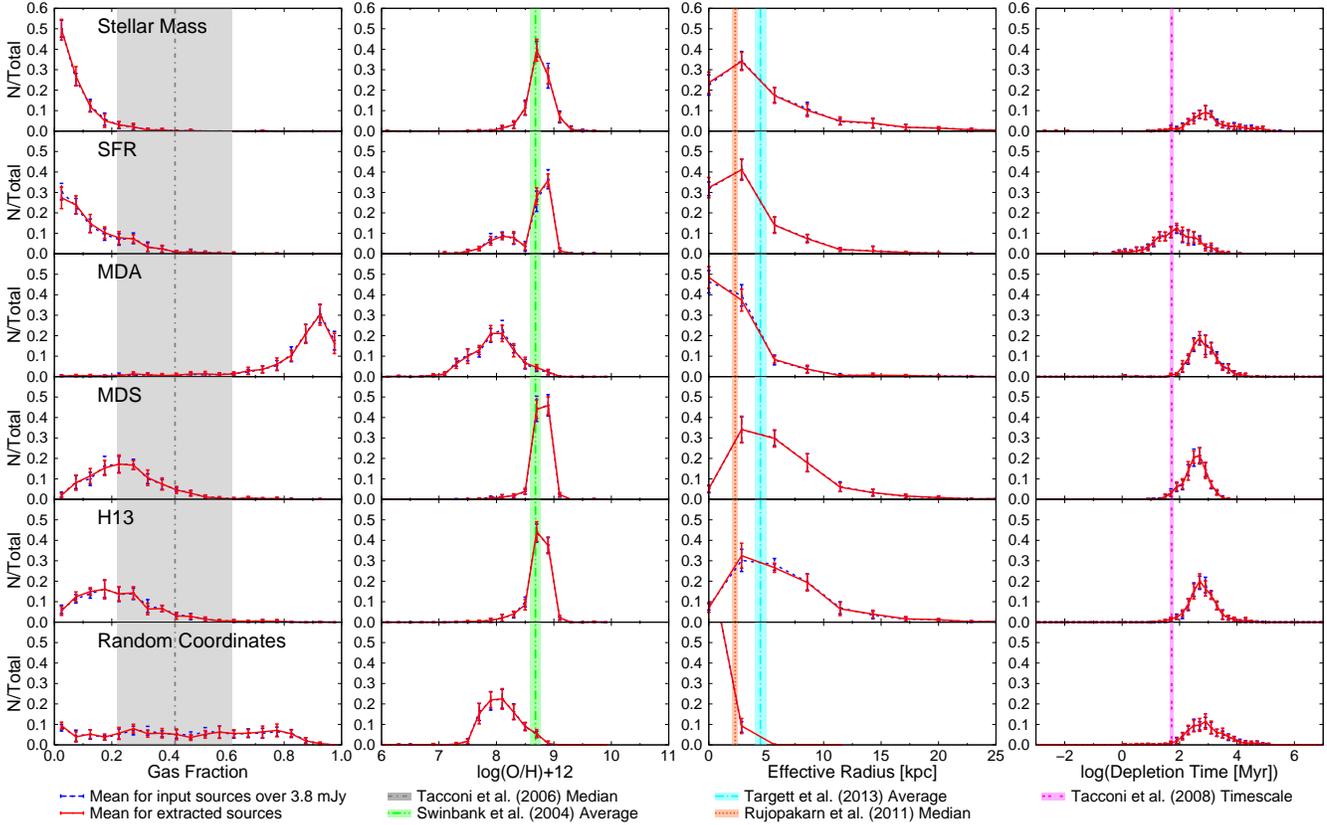}
\caption{Mean distributions over the ten lightcones of several properties obtained for five selected proxies: stellar mass, SFR, dust mass divided by stellar age (MDA), dust mass times SFR (MDS), and eq. 15 of \citet{Hayward2013} (H13), with red solid and blue dashed lines as in Fig. \ref{distribs1}. In columns from left to right, the predicted properties are: gas fraction, defined as cold gas mass over the sum between cold gas and stellar mass, compared to the median value found by \citet{Tacconi2006} as a vertical gray dot-dashed line; cold gas phase metallicity in terms of oxygen and hydrogen abundances, compared to the average value found by \citet{Swinbank2004} as a vertical green dot-dot-dashed line; effective radius, computed considering the contribution of both bulge and disc, compared with values found by \citet{Targett2013} and \citet{Rujopakarn2011} in vertical cyan long-dashed-dotted and orange dotted lines respectively; and depletion time, defined as the ratio between the cold gas mass and SFR, compared to the gas exhaustion timescale found by \citet{Tacconi2008} as a vertical magenta double-dotted line. Bottom panels show the distributions recovered when a random assignment of submm luminosities is done for model sources, injected in a map following a random spatial distribution.}
\label{distribs2}
\end{center}
\end{figure*}

\subsection{Prediction of Physical Properties of SMGs}

\begin{table*}
\begin{minipage}{\textwidth}
\begin{center}
\caption{Properties of sources extracted from the simulated maps. Uncertainties for each proxy are standard deviations of the median, considering all sources belonging to the 10 lightcones altogether.}
\begin{tabular}{ccccc}
Proxy & Redshift & $\log(\textmd{Stellar Mass}\,[\textmd{M}_{\odot}/h])$ & $\log(\textmd{Virial Halo Mass}\,[\textmd{M}_{\odot}/h])$ & SFR $[\textmd{M}_{\odot}/\textmd{yr}]$\\ \hline \hline
Stellar Mass & $0.223\pm0.261$ & $11.135\pm0.764$ & $12.716\pm1.353$ & $1.552\pm3.406$\\
SFR & $2.542\pm2.333$ & $11.074\pm0.552$ & $12.447\pm0.785$ & $287.299\pm173.629$\\
SFR Surface Density & $4.715\pm2.922$ & $9.972\pm0.979$ & $10.959\pm0.809$ & $61.755\pm61.321$\\
sSFR & $7.923\pm1.251$ & $7.982\pm0.208$ & $9.778\pm0.661$ & $4.540\pm1.984$\\
Dust Mass & $0.247\pm0.437$ & $10.993\pm0.977$ & $12.566\pm1.237$ & $8.239\pm18.077$\\
MDA & $3.656\pm2.759$ & $8.716\pm1.081$ & $10.996\pm1.238$ & $10.105\pm15.559$\\
MDD & $4.677\pm3.114$ & $10.392\pm0.645$ & $11.362\pm0.581$ & $102.731\pm61.160$\\
MDS & $1.862\pm1.448$ & $11.186\pm0.463$ & $12.728\pm0.498$ & $190.734\pm160.044$\\
H13 & $0.998\pm1.785$ & $11.177\pm0.528$ & $12.674\pm0.696$ & $129.016\pm237.874$\\
\hline
\end{tabular}
\label{medprops}
\end{center}
\end{minipage}
\end{table*}

\begin{table*}
\begin{minipage}{\textwidth}
\begin{center}
\caption{Properties of sources extracted from the simulated maps (cont.).}
\begin{tabular}{cccccc}
Proxy & sSFR $[\textmd{Gyr}^{-1}]$ & Gas Fraction & log(O/H)+12 & Effective Radius $[\textmd{kpc}]$ & Depletion Time $[\textmd{Myr}]$\\ \hline \hline
Stellar Mass & $0.014\pm0.030$ & $0.050\pm0.077$ & $8.767\pm0.220$ & $3.422\pm5.238$ & $5035.675\pm10884.577$\\
SFR & $1.652\pm2.242$ & $0.097\pm0.144$ & $8.769\pm0.222$ & $2.199\pm3.074$ & $ 70.547\pm130.123$\\
SFR Surface Density & $4.787\pm5.569$ & $0.007\pm0.008$ & $8.288\pm0.401$ & $0.021\pm0.012$ & $1.617\pm1.151$\\
sSFR & $ 35.106\pm7.081$ & $0.762\pm0.158$ & $7.834\pm0.141$ & $0.153\pm0.164$ & $101.953\pm94.529$\\
Dust Mass & $0.115\pm0.246$ & $0.215\pm0.201$ & $8.671\pm0.319$ & $6.940\pm8.190$ & $2260.248\pm3800.487$\\
MDA & $ 12.474\pm12.398$ & $0.896\pm0.113$ & $7.975\pm0.591$ & $1.479\pm1.953$ & $628.211\pm826.719$\\
MDD & $3.197\pm3.538$ & $0.009\pm0.014$ & $8.476\pm0.258$ & $0.039\pm0.036$ & $2.938\pm3.372$\\
MDS & $0.878\pm1.061$ & $0.238\pm0.172$ & $8.796\pm0.134$ & $5.127\pm4.998$ & $377.963\pm429.295$\\
H13 & $0.429\pm0.705$ & $0.203\pm0.181$ & $8.773\pm0.168$ & $5.222\pm5.838$ & $576.980\pm738.643$\\
\hline
\end{tabular}
\label{medprops2}
\end{center}
\end{minipage}
\end{table*}

Our analysis of the submm extracted sources is also extensible to other galaxy properties given by the semi-analytic model. Figs. \ref{distribs1} and \ref{distribs2} show some of the distributions predicted when taking $k$-corrected galaxies at $z>0$ in the count matching process: stellar mass, halo mass, SFR, sSFR, gas fraction, cold gas phase metallicity, effective radius and depletion time. Note that the intrinsic values of these properties are not affected by the count matching, since they are computed entirely by the \textsc{SAG} model. Keeping the values of these properties without modifications, assures that the final \textsc{SAG} galaxy population remains unaffected by the count matching technique. What may vary across the proposed proxies are the distributions of these properties for bright SMGs, since model sources having assigned the highest $870\,\mu\textmd{m}$ flux densities for one proxy could have, for instance, intrinsically lower SFRs when compared to model bright sources selected using another proxy.

All the properties listed above can be compared with observations\footnote{When comparing results for the proxies with observational properties that depend on assumed evolutionary synthesis models, we do not make any rescaling for IMFs different from the one assumed in this work (Salpeter IMF).}. For brevity, this is presented only for five of the proxies; we select the proxies giving the best agreement with the observed redshift distribution shown in Fig. \ref{zdistribinp}, as well as some of the proxies that do not. This is done in order to check if the former ones predict, in a consistent way, other galaxy properties when compared to observations. Median values for these properties are shown in Tables \ref{medprops} and \ref{medprops2}. The Arp220 template is used for $k$-correcting the galaxy spectra. Again, there is no significant difference between the recovered distributions and those for the input brightest sources.

We compare the distribution of stellar masses with the median values derived by \citet{Wardlow2011} and \citet{Michalowski2010}, who consider single-dish detected sources mapping the submm continuum; the former analyze 78 detected counterparts to 72 SMGs in the ECDF-S, while the latter consider 76 SMGs from the \citet{Chapman2005} sample. Except for the MDA proxy, all of them give reasonable distributions compared to SMG observations; note that the standard errors reported for observational quantities do not take into account the individual uncertainties in their determination. However, moving to the predicted virial halo masses, we find that only MDS and H13 proxies are able to reproduce the value estimated by \citet{Hickox2012} for SMGs in the ECDF-S field.

Our model underpredicts the SFRs by a factor of $\sim3$ when compared to observed values for SMGs by \citet{Michalowski2010}, being more consistent with SFRs measured for high-$z$ star-forming galaxies (SFGs) by \citet{Tacconi2010}, which have mean values of 95 and $135\,\textmd{M}_{\odot}/\textmd{yr}$ at $z=1.2$ and 2.3 respectively (note that this is not a sample of SMGs). Deblending FIR observations for the ALESS SMG positions with \textit{Herschel}, \citet{Swinbank2014} find a median SFR $\sim2$ times lower than the one derived for single-dish selected SMGs, although their sample includes $870\,\mu\textmd{m}$ fluxes down to $2\,\textmd{mJy}$ (including only sources brighter than $4.2\,\textmd{mJy}$ gives an SFR of $530\pm60\,\textmd{M}_{\odot}/\textmd{yr}$). Regarding the sSFR distribution, the stellar mass proxy gives considerably lower values compared to the median value found by \citet{Michalowski2010}, the MDA proxy considerably higher values, and the H13 proxy a bimodal distribution; the remaining proxies give a reasonable prediction.

Reproducing the observed SFR distribution for SMGs is an outstanding challenge for semi-analytic models. Some aspects in \textsc{SAG} regarding the modeling of the star formation (e.g. the complete removal of hot gas when galaxies become satellites, or not including cold gas inflows) might be leading to lower predicted SFRs not only for SMGs, but also for the whole galaxy population across redshift (see the discussion about Fig. \ref{sagsfrvol} in Appendix \ref{sagpred}). This translates in globally underpredicting galaxy SFRs at the redshifts where SMGs lie. It affects the predicted distributions for SMGs given by all our proposed proxies, recalling that the intrinsic SFR of each galaxy is not affected by the count matching process.

However, for each model SMG we can explore the relation between the SFR, given by the semi-analytic model, and the bolometric IR luminosity, computed from its $870\,\mu\textmd{m}$ flux (using its redshift and the SED template to integrate its emission in the rest-frame wavelength range $8-1000\,\mu\textmd{m}$). The MDS proxy predicts a positive trend between SFR and bolometric IR luminosity, in line with observations: on average, a model source classified as ULIRG has larger SFR than one classified as LIRG. We highlight this prediction because, when performing the count matching, we pose no requirements regarding this relationship. Therefore, different proxies can give completely different SFR vs bolometric IR luminosity laws.

We also compute the model gas fraction for the brightest submm sources, as the ratio between the cold gas mass and the sum of cold gas and stellar mass. Compared to the average that \citet{Tacconi2006} found for 8 SMGs at $z\sim2-3.4$ (estimated from both continuum and CO emission lines), the MDS and H13 proxies give less gas-rich sources than the observed SMGs, while the stellar mass and SFR proxies give very gas-poor galaxies, and the MDA proxy very gas-rich sources. In all cases except for MDA proxy (where it is underpredicted), the metallicity of the remaining cold gas in the galaxy is close to the average value that \citet{Swinbank2004} obtained for 15 sources including SMGs and optical faint radio galaxies (OFRGs, \citealt{Chapman2004}) targeting the $\textmd{H}\alpha$ line.

We compute the effective (i.e., projected) radius of a given galaxy weighting the contributions of both bulge and disc as
\begin{equation}
r_{\textmd{eff}}=\frac{M_{\textmd{bulge}}\,r_{\textmd{bulge,eff}}+M_{\textmd{disc}}\,r_{\textmd{disc,eff}}}{M_{\textmd{bulge}}+M_{\textmd{disc}}}, \label{}
\end{equation}
\noindent where $M_{\textmd{bulge}}$ is the bulge stellar mass, $M_{\textmd{disc}}$ is the sum of cold gas and stellar mass of the disc, $r_{\textmd{bulge,eff}}=r_{\textmd{bulge}}/1.35$ and $r_{\textmd{disc,eff}}=r_{\textmd{disc}}/1.68$ are the effective radii for bulge\footnote{A brief description of the adopted model for bulge sizes is given in Appendix \ref{appbulges}.} and disc respectively, given in terms of the half-mass radii in three dimensions; the factors 1.35 and 1.68 correspond to concentration values for elliptical and spiral galaxies respectively, taken from \citet{Graham2005}. Compared to the average half-light radius found by \citet{Targett2013} for 24 SMGs in the GOODS-S and to the median effective radius found by \citet{Rujopakarn2011} for 48 sources at high redshift (median of 1) comprising SMGs and ULIRGs, we find that the stellar mass and SFR proxies give distributions peaking at the value for observed high-$z$ ULIRGs and SMGs, while the MDA proxy isolates mostly very compact sources. The MDS and H13 proxy prefer sources with size in agreement with \citet{Targett2013} data, although both distributions are quite broad.

\begin{figure*}
\begin{center}
\includegraphics[width=0.6\textwidth]{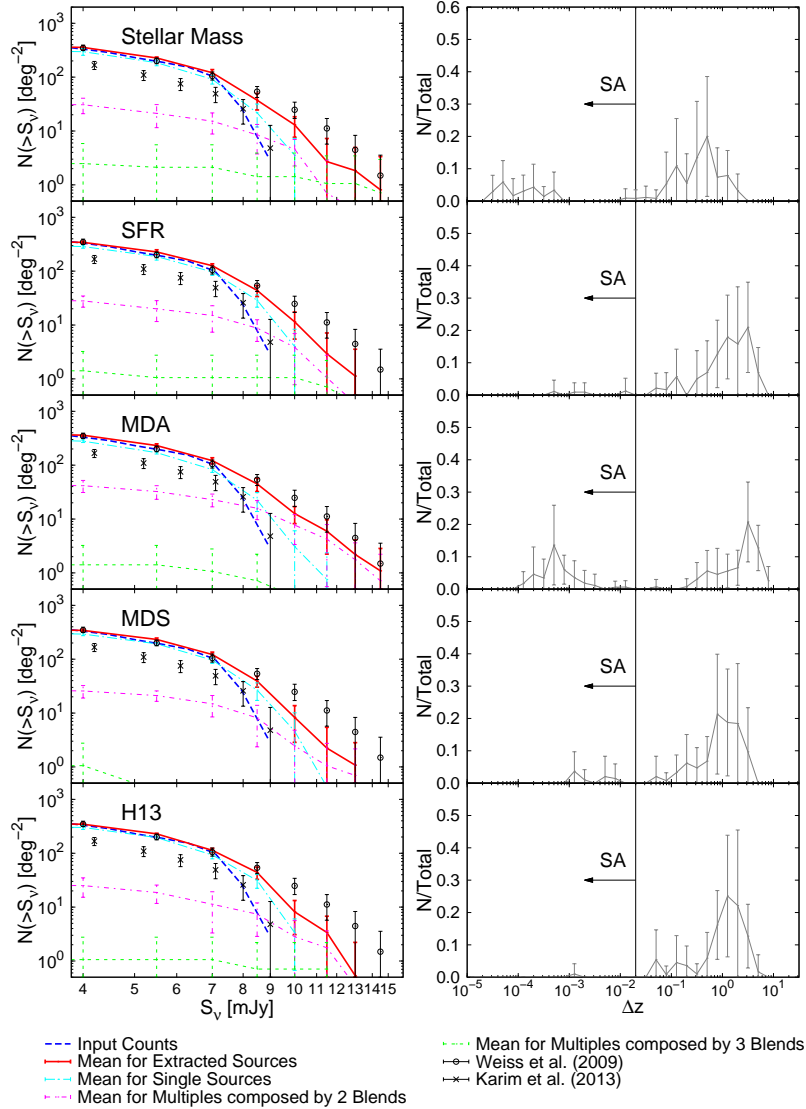}
\caption{Properties of multiple sources for five selected proxies. Left column: recovered cumulative number counts at $870\,\mu\textmd{m}$ (red thick solid lines) split in the contribution by single sources (cyan thin long-dashed-dotted lines), multiples composed by two blended sources (magenta thin dot-dashed lines) and multiples composed by three blended sources (green thin short-dashed lines); these are compared with the input counts (blue thick dashed lines) and to observational data from LABOCA and ALMA studies (symbols as in Fig. \ref{countscones}). Right column: mean distribution of the redshift separation between components of multiple sources (gray solid lines). ``SA'' stands for spatially associated sources, according to the cut proposed by \citet{Hayward2013b}.}
\label{countsblended_deltaz}
\end{center}
\end{figure*}

Finally, we compute the depletion time for model galaxies, defined as the ratio between the cold gas mass and SFR. This gives a measure of the gas exhaustion timescale. Only the SFR proxy gives a distribution consistent with the \citet{Tacconi2008} timescale for 4 SMGs, which was derived from mm CO interferometry. Conversely, the MDA, MDS and H13 proxies give distributions consistent with the value found by \citet{Tacconi2010} for high-$z$ SFGs ($0.9\pm0.6\,\textmd{Gyr}$); the stellar mass proxy gives a similar distribution too, but as is shown in its SFR distribution, most of the brightest galaxies have SFRs close to zero, giving depletion times tending to infinite values, which do not appear in the depletion time plot but are still considered when calculating the median value for the model (see Table \ref{medprops2}).

As was done for galaxy redshifts, we compute the recovered distribution of all these properties when a random assignment of submm luminosities is done for model sources, being injected in the simulated map with a random spatial distribution. This is shown in the bottom panels of Figs. \ref{distribs1} and \ref{distribs2}. The distributions obtained in this case are quite different from those for observed SMGs, except for the sSFRs, where the apparent agreement with observations comes from the distribution of this property for the whole model galaxy population.

It is worth noting that the validity of the predicted properties is preserved even after taking into account scatter in the proxy, as well as when different SED templates (\citealt{Michalowski2010} or \citealt{Magdis2012}) are assumed to compute $k$-corrections. Nevertheless, when excluding AGNs in the \textsc{SAG} model, SFR distributions peak at lower values for all proxies. In particular, it decreases the median SFR by $\sim200\,\textmd{M}_{\odot}/\textmd{yr}$ for the MDS proxy. Predicted stellar masses change slightly, remaining consistent with observational data. The large changes are expected though, as this model is extremely different from the fiducial one; the bright end of the local optical luminosity function is affected, as well as the slope of the cosmic SFR across redshift, departing considerably from observations. On the other hand, adopting the \citet{DeLucia2004} prescription for star formation gives roughly the same stellar and virial masses compared to the fiducial case, except for the H13 proxy, where a tail with lower masses appears. For the SFR and MDS proxies the SFRs are larger, changing the median value by $\sim50\,\textmd{M}_{\odot}/\textmd{yr}$ (affecting their depletion times accordingly). The largest impact is on the gas fractions, which now peak at 0 (recall that in this prescription stars can form for any amount of cold gas). For the stellar mass proxy, a tail with large metallicities appears.

Therefore, models with very different star formation laws still broadly agree in the predicted SMG properties, unlike models where AGNs are not taken into account. As long as the model provides a reasonable fit to the general galaxy population, the predictions are robust.

\subsection{Effect of Galaxy Blending on the Recovered Counts}

We study the contribution of multiple sources to the cumulative number counts at submm wavebands, separating them from single sources. The recovered fraction of blends in a multiple source depends naturally on the flux depth, as is pointed out by \citet{Chen2013}. At the counts level shown in Fig. \ref{countsblended_deltaz}, left column, we obtain a maximum of three blends down to an input flux level of $1.2\,\textmd{mJy}$. For the stellar mass, SFR and H13 proxies, there are no single sources brighter than $10\,\textmd{mJy}$, but only multiples. For MDA and MDS proxies, the importance of multiple sources composed by two blends begins at extracted fluxes higher than $\sim9$ and $\sim11\,\textmd{mJy}$ respectively.

\begin{figure}
\begin{center}
\includegraphics[width=0.41\textwidth]{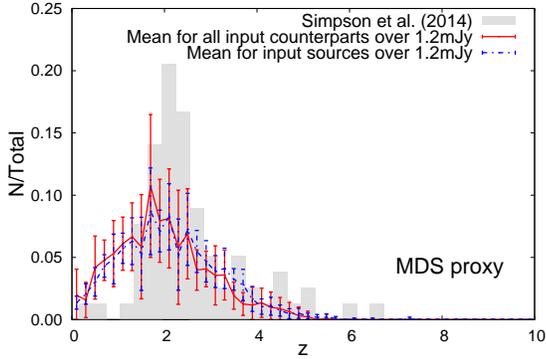}
\caption{Mean redshift distribution over the ten lightcones for the MDS proxy, considering all the counterparts for extracted sources in the input catalog over $1.2\,\textmd{mJy}$ (red solid line). This is compared with the distribution of photometric redshifts for ALMA sources in the ECDF-S from \citet{Simpson2014} (gray histogram). The mean distribution for all input sources over $1.2\,\textmd{mJy}$, i.e., not only around bright extracted sources, is also shown (blue dot-dashed line). Standard deviations are reported for each redshift bin.}
\label{hist_z_match_input_mds}
\end{center}
\end{figure}

\begin{figure}
\begin{center}
\includegraphics[width=0.35\textwidth]{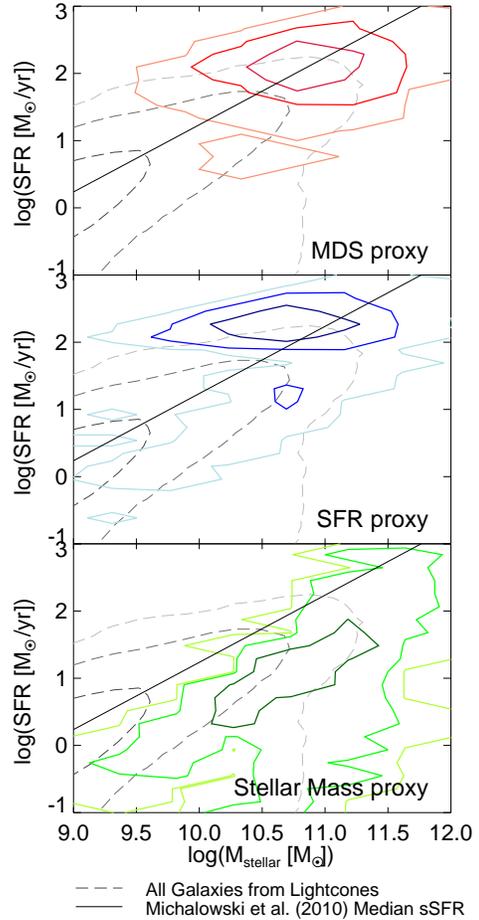}
\caption{Star formation rate versus stellar mass including sources extracted from all simulated maps. Three proxies are compared in separate diagrams for clarity: the results for our best proxy, MDS (red solid contours) are compared to those for stellar mass and SFR proxies (green and blue solid contours respectively). Also displayed are the distribution for galaxies in the ten input lightcones (gray dashed contours), and the relation between SFR and stellar mass derived from the median specific SFR determined by \citet{Michalowski2010} (black solid line). From the darkest to the lightest, contours enclose 68, 95 and 99 per cent of the sources.}
\label{sfr_mst}
\end{center}
\end{figure}

In order to address if there is a spatial association between the blended sources that compose a multiple one, we compute the redshift separation between components as
\begin{equation}
\Delta z=\sqrt{\sum_{i\neq j}(z_i-z_j)^2}. \label{deltazeq}
\end{equation}
The mean distributions are shown in Fig. \ref{countsblended_deltaz}, right column. For distinguishing between associated and unassociated sources, we follow the limit of $\Delta z=0.02$ proposed by \citet{Hayward2013b}. For the five proxies we find that most of the components are unassociated (in agreement with findings by \citealt{Hayward2013b} and \citealt{Cowley2014} for model SMGs), with a significant amount of associated sources only for the stellar mass and MDA proxies. This result is in line with the lack of a dependence in the counts on the clustering of sources of different proxies, which is also similar to the case of random positions (see Section \ref{counts}); in a given submm image, most sources that lie in a same line of sight are there by chance.

\subsection{Characteristics of our Best Proxy: MDS}

We end this section presenting some predictions for SMGs in the MDS proxy, which is the only one giving at the same time redshift, stellar mass and host halo mass distributions consistent with observed ones for SMGs.

\subsubsection{Multiplicity of Single-Dish Detected SMGs and Redshift Distribution of all Blends}

\begin{table*}
\begin{center}
\caption{Mean cumulative fraction over flux density $S_{\nu}$ at $870\,\mu\textmd{m}$ for MDS proxy. Standard deviations are reported.}
\begin{tabular}{c|cccc}
Type of & \multicolumn{4}{c}{$N/N_{\textmd{extractions}}(>S_{\nu})$}\\ \cline{2-5}
source & $7\,\textmd{mJy}$ & $8.5\,\textmd{mJy}$ & $10\,\textmd{mJy}$ & $11.5\,\textmd{mJy}$\\ \hline \hline
Single & $0.842\pm0.134$ & $0.714\pm0.308$ & $0.584\pm0.795$ & $0.199\pm0.690$\\ \hline
Multiple composed & \multirow{2}{*}{$0.155\pm0.068$} & \multirow{2}{*}{$0.274\pm0.206$} & \multirow{2}{*}{$0.363\pm0.419$} & \multirow{2}{*}{$0.595\pm1.275$}\\
by 2 blends & \\ \hline
Multiple composed & \multirow{2}{*}{$0.003\pm0.012$} & \multirow{2}{*}{$0.012\pm0.039$} & \multirow{2}{*}{$0.053\pm0.173$} & \multirow{2}{*}{$0.206\pm0.712$}\\
by 3 blends & \\ \hline \hline
Total & 1 & 1 & 1 & 1\\ \hline
\end{tabular}
\label{fracblended}
\end{center}
\end{table*}

\begin{figure*}
\begin{center}
\includegraphics[width=0.84\textwidth]{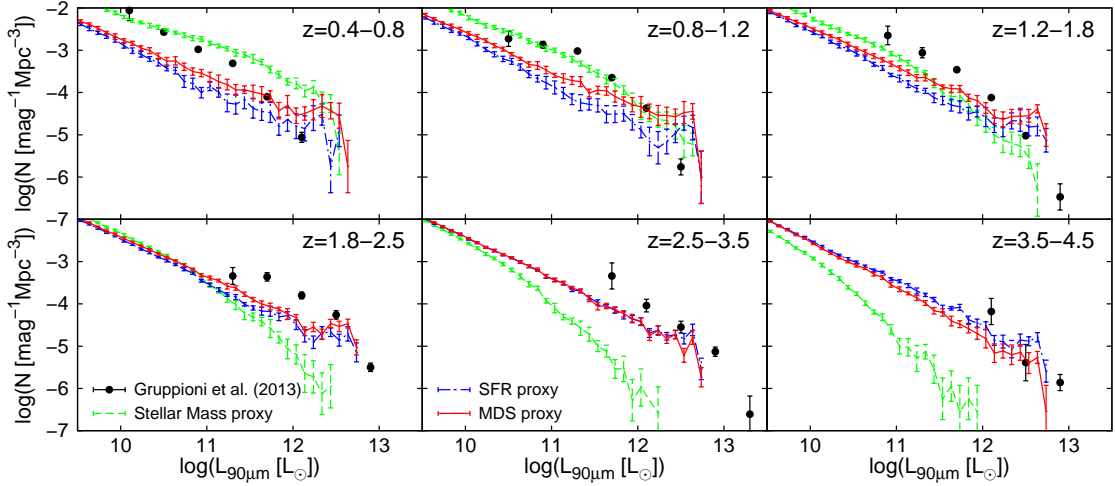}
\caption{Rest-frame luminosity functions at $90\,\mu\textmd{m}$ predicted for the brightest input sources in each proxy, at six redshift ranges. Results are presented for MDS (red solid lines), stellar mass (green dashed lines) and SFR proxies (blue dot-dashed lines), averaging over the ten lightcones. Standard deviations are reported. Observational data by \citet{Gruppioni2013} of \textit{Herschel} sources in the Guaranteed Time Observation PACS Evolutionary Probe Survey are also shown (black filled circles).}
\label{lf_match}
\end{center}
\end{figure*}

Table \ref{fracblended} shows the cumulative fraction $N/N_{\textmd{extractions}}(>S_{\nu})$ of single and multiple sources at several submm fluxes, averaged over the ten lightcones. Note that over $8.5\,\textmd{mJy}$ this proxy predicts that $\sim27$ per cent of the detected sources are multiples composed by two blends. Over $10\,\textmd{mJy}$ the statistics is noisier, but still not consistent with the finding by \citet{Karim2013} of $100$ per cent multiple sources over $\sim9\,\textmd{mJy}$ in the ECDF-S. Over our S/N limit of $3.8$, the fraction of multiples is in the range $\sim9-14$ per cent (see Table \ref{10cones}). This is lower than the range reported by \citet{Hodge2013} for LESS sources, $\sim35-45$ per cent. Conversely, it is closer to the findings of $12.5_{-6.8}^{+12.1}$ per cent by \citet{Chen2013} for SCUBA-2 SMGs in the CDF-N, and to the $22\pm9$ per cent by \citet{Smolcic2012} for LABOCA SMGs in the COSMOS field.

We explore whether the redshift distribution of both single sources and the blends of multiple sources is different from the distribution of all model sources over our flux cut of $1.2\,\textmd{mJy}$  in the whole field. This statistics could be obtained for the actual ECDF-S once interferometric observations over the entire field are carried out. So far, the determination of photometric redshifts for actual sources over $\sim1.2\,\textmd{mJy}$ in the ECDF-S is restricted to ALMA sources surrounding/comprising bright LABOCA SMGs \citep{Simpson2014}; they consider a subset of the \citet{Hodge2013} catalog.

We show the comparison between model and observed data in Fig. \ref{hist_z_match_input_mds}. Compared to ALMA sources, in proportion we recover more sources at low redshift. Restricting the discussion to our model data, there is a slight change in the distribution taking input sources in the whole field compared to the subset of sources that lie around or comprise SMGs extracted from the simulated maps: considering all sources in the field, we recover slightly less low-$z$ and more high-$z$ sources in proportion.

\subsubsection{Star Formation Rate vs Stellar Mass}

We also compare our best proxy with the stellar mass and SFR ones, putting together the sources selected by each proxy in all lightcones, after going through the observational process. For each source extracted from the simulated maps, we consider the properties of all counterparts within the search radius down to $1.2\,\textmd{mJy}$.

Fig. \ref{sfr_mst} shows the SFR as a function of stellar mass for the three proxies, including also the distribution for all galaxies in the ten input lightcones and the median sSFR for SMGs by \citet{Michalowski2010}, translated to a function with the shape $\textmd{SFR}\propto\textmd{M}_{\textmd{stellar}}$. Results are in line with the findings shown in Fig. \ref{distribs1}, fourth column; the MDS and SFR proxies give most of the SMGs lying in the high-mass end of the relationship between SFR and stellar mass that can be derived from the observational median sSFR, while the stellar mass proxy gives SMGs lying below the relation.

\subsubsection{Luminosity Functions}

Rest-frame luminosity functions (LFs) at other wavebands can be recovered using the input $870\,\mu\textmd{m}$ flux and redshift of all sources in a given lightcone and making use of the Arp220 template spectrum. For instance, Fig. \ref{lf_match} shows rest-frame LFs at $90\,\mu\textmd{m}$ in six redshift ranges spanning from $z=0.4$ to 4.5, predicted for model SMGs in each of the three proxies. The MDS and SFR proxies predict similar distributions at all redshifts, giving a reasonable agreement with \citet{Gruppioni2013} \textit{Herschel} data at $z=3.5-4.5$. For all these proxies, the underprediction of the observed LFs at $z=1.8-3.5$ is related to overpredicting low-$z$ ones; this is particularly notorious for the stellar mass proxy even at the highest redshifts. Trying different SED templates for computing $k$-corrections, as described in Section \ref{zdistrib}, does not improve our results.

In this case, the observational distributions are presented only as a reference, since in principle we do not expect to recover similar LFs. \citet{Gruppioni2013} use samples from the Guaranteed Time Observation PACS Evolutionary Probe Survey, selected through their emission observed at $160\,\mu\textmd{m}$, whereas in our case only the $870\,\mu\textmd{m}$ waveband is used as a selection criterion. In addition, our assumptions of fixed spectra may not represent galaxies at all the redshifts with which we are making the comparison.

\subsubsection{Descendants of Model SMGs}

Once we determine the input counterparts for sources extracted from simulated maps, we follow their history in the \textsc{SAG} model until the present epoch.

\begin{figure}
\begin{center}
\includegraphics[width=0.34\textwidth]{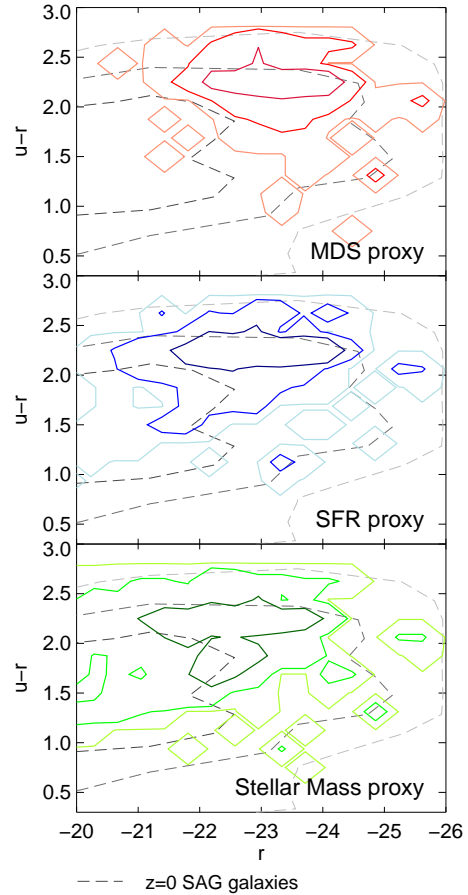}
\caption{Colour magnitude diagram for descendants at $z=0$ of extracted sources from all simulated maps. Magnitudes are in the rest-frame. Same proxies as in Fig. \ref{sfr_mst} are considered, now compared to the distribution for all $z=0$ \textsc{SAG} galaxies having stellar masses over $10^8\,\textmd{M}_{\odot}$ (gray dashed contours). From the darkest to the lightest, contours enclose 68, 95 and 99 per cent of the sources.}
\label{des_z0}
\end{center}
\end{figure}

A colour magnitude diagram for these descendants is shown in Fig. \ref{des_z0} for the three proxies, giving also the distribution for all $z=0$ \textsc{SAG} galaxies with stellar mass over $10^8\,\textmd{M}_{\odot}$. While the stellar mass proxy gives a significant population of blue-cloud galaxies, descendants for the MDS and SFR proxies lie in the red-sequence region. Note that in this approach the history of model $z=0$ early-type galaxies is not followed backwards, so this does not necessarily ensures that the SMG phase is common to all early-type galaxies, nor that passing through an SMG phase is the only way to build up a red-sequence galaxy in our model.

Fig. \ref{des_z0distribs} shows the distribution of stellar mass, virial halo mass, SFR and galaxy type for $z=0$ descendants given by each proxy. As a reference, we also present observational data for the Milky Way: the stellar mass range computed by \citet{Flynn2006} using the local optical LF and the vertical structure of the Galaxy disc, the total mass within $\sim200\,\textmd{kpc}$ obtained by \citet{Bhattacharjee2014} measuring the rotation curve of the Galaxy, and the SFR determined by \citet{Murray2010} using the total Galaxy free-free emission.

While in all cases most of the descendants are central galaxies having SFRs in the range $0-10\,\textmd{M}_{\odot}/\textmd{yr}$, the distribution of stellar masses is different for each proxy, being the lowest for the stellar mass proxy and the highest for the MDS one. Regarding the virial halo masses, the stellar mass proxy gives a very wide distribution, while the SFR and MDS proxies give mainly sources lying at and to higher values than the median $z=0$ value predicted by \citet{Hickox2012} of $\log(M_{\textmd{vir}}\,[\textmd{M}_{\odot}/h])=13.3_{-0.5}^{+0.3}$. Comparing the SFR distribution for descendants with the one for model $z=0$ ellipticals, or even extending the comparison to all model $z=0$ galaxies (see the inset in Fig. \ref{des_z0distribs}), the average descendant of SMG tends to have a higher tail of high SFR; this is found regardless of the proxy. At this point it is worth noting the presence of outliers in the large sample of \textsc{SAG} model galaxies used to construct the lightcones. Among these are some sources with red colours but SFRs over $\sim10\,\textmd{M}_{\odot}/\textmd{yr}$, which will leave the red sequence after a few tens of Myr.

\section{Summary}\label{summary}

\begin{figure*}
\begin{center}
\includegraphics[width=0.74\textwidth]{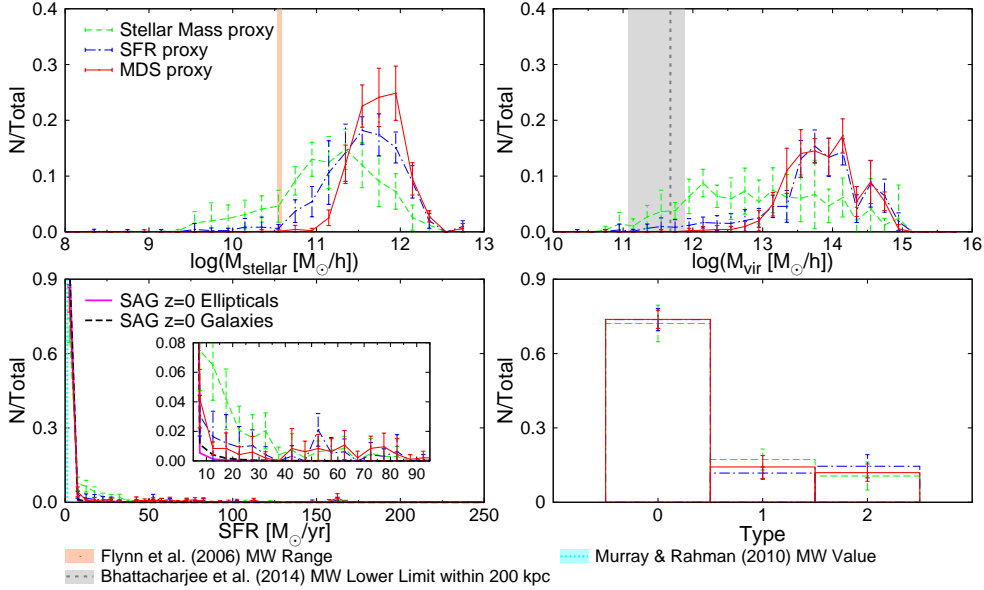}
\caption{Distributions of stellar mass, virial halo mass, SFR and galaxy type for descendants at $z=0$ of extracted sources. Histograms are averaged over the ten lightcones, taking the same proxies as in Fig. \ref{sfr_mst}. In the bottom left panel, the SFR distribution for model $z=0$ ellipticals and for all model $z=0$ galaxies are shown as magenta thick solid and black thick dashed lines respectively. Galaxy types shown in the bottom right panel correspond to: 0 for central galaxies, 1 for substructure satellites and 2 for satellites whose substructure has been lost, but which have not yet merged. As a reference, observational data for the Milky Way (MW) are shown: the stellar mass range computed by \citet{Flynn2006} as a light orange region, the total mass within $\sim200\,\textmd{kpc}$ obtained by \citet{Bhattacharjee2014} as a vertical gray short-dashed line, and the SFR determined by \citet{Murray2010} as a vertical cyan dotted line. The errors for observational values are presented as coloured shaded regions.}
\label{des_z0distribs}
\end{center}
\end{figure*}

In light of the new findings regarding SMGs in the ECDF-S with ALMA, namely, the brightest single-dish detected sources being comprised by emission from multiple fainter sources \citep{Karim2013}, we develop a new technique for modeling SMGs, using lightcones of galaxies drawn from the semi-analytic model \textsc{SAG} in a $\Lambda$CDM framework. In this paper we introduce the count matching approach, where physical galaxy properties (or their combinations) given by the model are selected as proxies for the submm galaxy emission. Submm luminosities are assigned to a mass-limited sample of model galaxies assuming a monotonic relationship with the proxies, in such a way that a combination of LABOCA \citep{Weiss2009} plus bright-end ALMA observed number counts \citep{Karim2013} are reproduced. We make model catalogs given by each proxy pass through the observational process just as single-dish observations at $870\,\mu\textmd{m}$, allowing us to recover cumulative number counts as is done for actual data. Going back to the properties of counterparts for each extracted source (i.e., the injected sources whose coordinates are within a given search radius from each detection), we compare the recovered distributions of several galaxy properties with current observations from the literature.

Our main findings are the following:

\begin{itemize}
\item For the nine proposed proxies, there are lines of sight that give cumulative submm number counts consistent with LABOCA data \citep{Weiss2009}. This is found even for maps having the same distribution of injected fluxes but randomized source coordinates, giving a hint that the clustering has a minor influence in the counts (see Fig. \ref{countscones}).
\item The majority of components of blended SMGs are spatially unassociated, meaning that most sources lying in a single line of sight are there by chance (see Fig. \ref{countsblended_deltaz}). Across the proxies, the fraction of multiple sources goes from $\sim9$ to $\sim14$ per cent, underpredicting the actual fraction for SMGs in the ECDF-S of $\sim35-45$ per cent \citep{Hodge2013} but closer to the values found by \citet{Chen2013} in the CDF-N ($12.5_{-6.8}^{+12.1}$ per cent) and \citet{Smolcic2012} in the COSMOS field ($22\pm9$ per cent).
\item The different proxies retrieve a variety of distributions for a given galaxy property. We explore SMG distributions of redshift, stellar mass, host halo mass, SFR, sSFR, gas fraction, metallicity, effective radius and depletion time (see Figs. \ref{distribs1} and \ref{distribs2}, and Tables \ref{medprops} and \ref{medprops2}).
\item The \textsc{SAG} model underpredicts the individual measured SFRs by a factor of $\sim3$ when compared to actual SMG data from single-dish observations. This could be affected by some aspects in the modeling of star formation, like the complete removal of hot gas when galaxies become satellites.
\item While the proxy where the submm luminosity increases monotonically with the ratio between dust mass and stellar age (the MDA proxy) is slightly better than the others when reproducing LABOCA cumulative counts, it fails to reproduce stellar and host halo masses, as well as the gas fraction and other properties.
\item The recovered redshift, stellar mass and host halo mass distributions (among others) for model SMGs are consistent with observations, when the submm luminosity increases monotonically with the product between dust mass and SFR (the MDS proxy). In this proxy most of the SMG $z=0$ descendants are central galaxies with SFRs between 0 and $10\,\textmd{M}_{\odot}/\textmd{yr}$, with $u-r$ rest-frame colours lying in the red sequence (see Figs. \ref{des_z0} and \ref{des_z0distribs}).
\item Using a fixed Arp220 for all model galaxies we are able to reproduce the rest-frame luminosity function observed by \citet{Gruppioni2013} at $90\,\mu\textmd{m}$ in the redshift range $z=3.5-4.5$, but underpredicting it at $z=1.8-3.5$. This is related to overpredicting low-$z$ distributions (see Fig. \ref{lf_match}). In future work, computing LFs at other FIR wavebands could also be used to determine a more precise proxy, specially if we use counts from larger surveys.
\end{itemize}

An immediate benefit of using this technique to assign submm luminosities to big galaxy samples is the low computational cost, unlike the use of full radiative transfer including dust over each galaxy, which computes the FIR emission in a self-consistent way. Our phenomenological approach gives results consistent with the \citet{Hayward2013b} prediction regarding the lack of spatial association between blended galaxies, and we see a remarkable similarity between the predictions of the MDS proxy with the fitting function used in their paper to compute the submm flux, which was obtained using results of the \textsc{SUNRISE} dust radiative transfer code \citep{Jonsson2006} on hydrodynamical simulations (see \citealt{Hayward2013} for details). The high proportion of spatially unassociated sources has also a qualitative agreement with \citet{Cowley2014} results, where the galaxy properties (including submm flux) were obtained using a semi-analytic model plus a dust model that resembles the treatment performed by the \textsc{GRASIL} spectrophotometric code.

Further improvements in the count matching approach could consider the enhancement in mass resolution for the dark matter simulation on which the semi-analytic model is run, the increase in the simulation volume, the incorporation of AGN properties as proposed proxies and the inclusion of a more gradual star formation during starbursts \citep{Gargiulo2014} or cold gas inflows in the \textsc{SAG} model. When modeling the observational process, it is also possible to quantify the influence of gravitational lensing on the fluxes and redshift distribution of detected sources, including the boosting of higher redshift sources above the survey limit; nevertheless, lensing would affect number counts well beyond our covered range for this field and wavelength, according to models (\citealt{Lima2010}, \citealt{Er2013}).

The flexibility in the methodology underlying the count matching approach can be extended to other galaxy populations and/or wavelength ranges and/or observed fields to be modeled, as well as to other galaxy formation models and/or cosmology. 

\section*{Acknowledgements}

We thank the referee for the comments and suggestions which contributed to improve this paper. We also acknowledge G. Bruzual and R. Hickox for kindly providing data and comments. AMMA acknowledges support from CONICYT Doctoral Fellowship program. This work was supported in part by Fondecyt Regular No. 1110328, BASAL PFB-06 ``Centro de Astronom\'ia y Tecnolog\'ias Afines''. FPN acknowledges support through the DFG CRC-956. NDP thanks the hospitality of the Max Planck Institute for Astrophysics at Garching, where part of this work was done and several helpful discussions were held. AMMA and NDP acknowledge support from the European Commissions Framework Programme 7, through the Marie Curie International Research Staff Exchange Scheme LACEGAL (PIRSES-GA-2010-269264). SAC acknowledges grants from CONICET (PIP-220), Agencia Nacional de Promoci\'on Cient\'ifica y Tecnol\'ogica (PICT-2008-0627), Argentina, and Fondecyt, Chile. EG and PK acknowledge support from the National Science Foundation through grant AST-1055919. The Geryon cluster at the Centro de Astro-Ingenier\'ia UC was extensively used for the calculations performed in this paper. The Anillo ACT-86, FONDEQUIP AIC-57, and QUIMAL 130008 provided funding for several improvements to the Geryon cluster.

\appendix

\section{Predictions from the \textsc{SAG} Model}\label{sagpred}

The semi-analytic model involves a set of free parameters that regulate the action of the different physical processes, namely gas cooling, star formation, feedback from core-collapse supernova explosions and AGN, galaxy mergers and chemical enrichment of baryons. The values of these parameters are tuned by using a calibration method based on the Particle Swarm Optimization technique \citep{Ruiz2014}. Observational constraints for this procedure are $z=0$ luminosity functions at $b_J$ and $r$ bands, and the black hole - bulge mass relation. Rest-frame magnitudes in several filters from the UV to the NIR are computed considering \textsc{CB07} evolutionary synthesis models (which are an update of \citealt{Bruzual2003} models) for different cold gas metallicities and assuming a Salpeter IMF. This magnitudes are corrected by dust extinction following \citet{DeLucia2004}. To validate the model used throughout this work, we present some of the predicted galaxy properties.

\begin{figure}
\begin{center}
\includegraphics[width=0.39\textwidth]{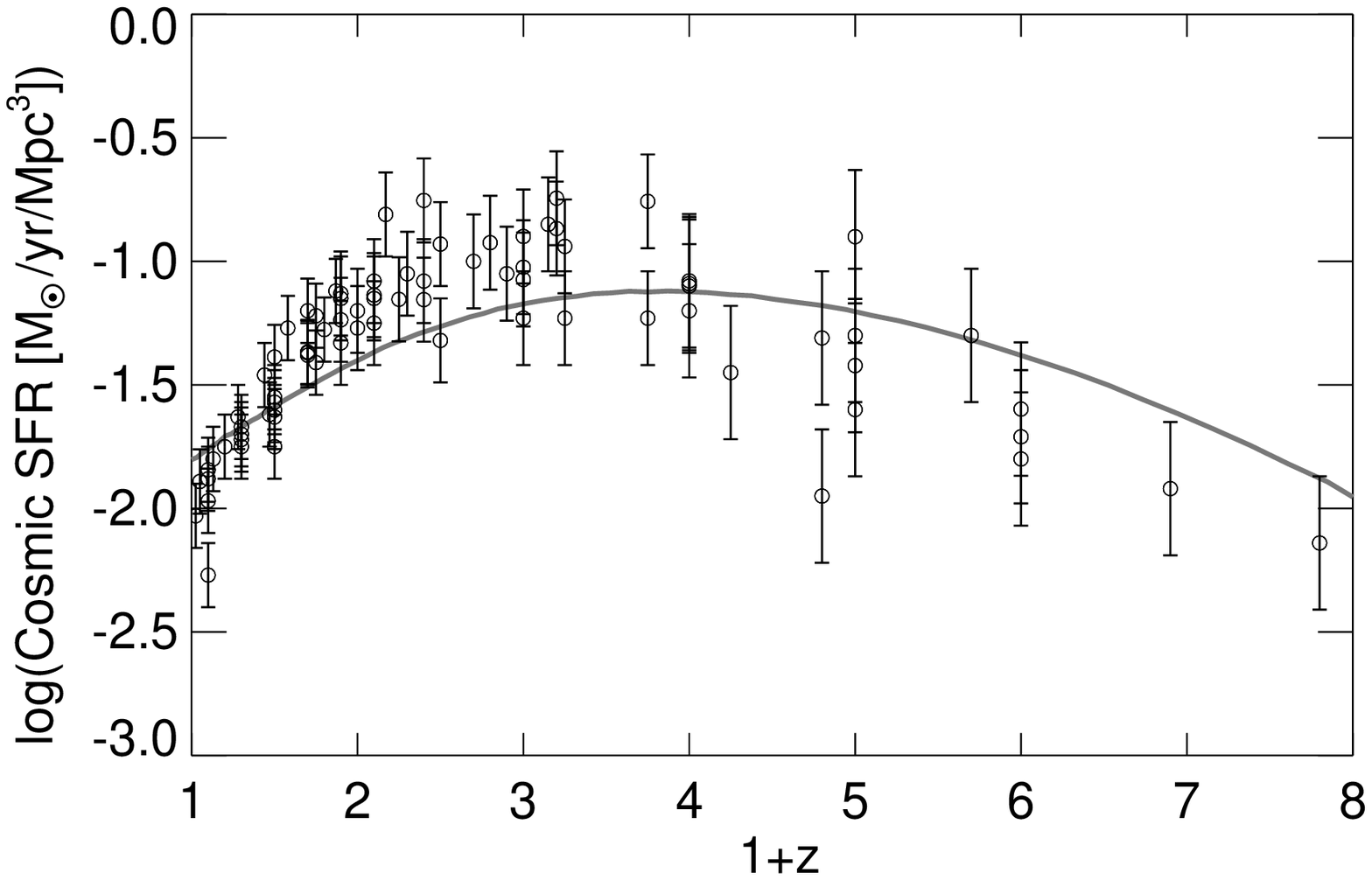}
\caption{Cosmic star formation rate as a funcion of redshift predicted by \textsc{SAG} (gray line) versus observational data compiled by \citet{Behroozi2013} (open circles).}
\label{sagsfrvol}
\end{center}
\end{figure}

\begin{figure}
\begin{center}
\includegraphics[width=0.37\textwidth]{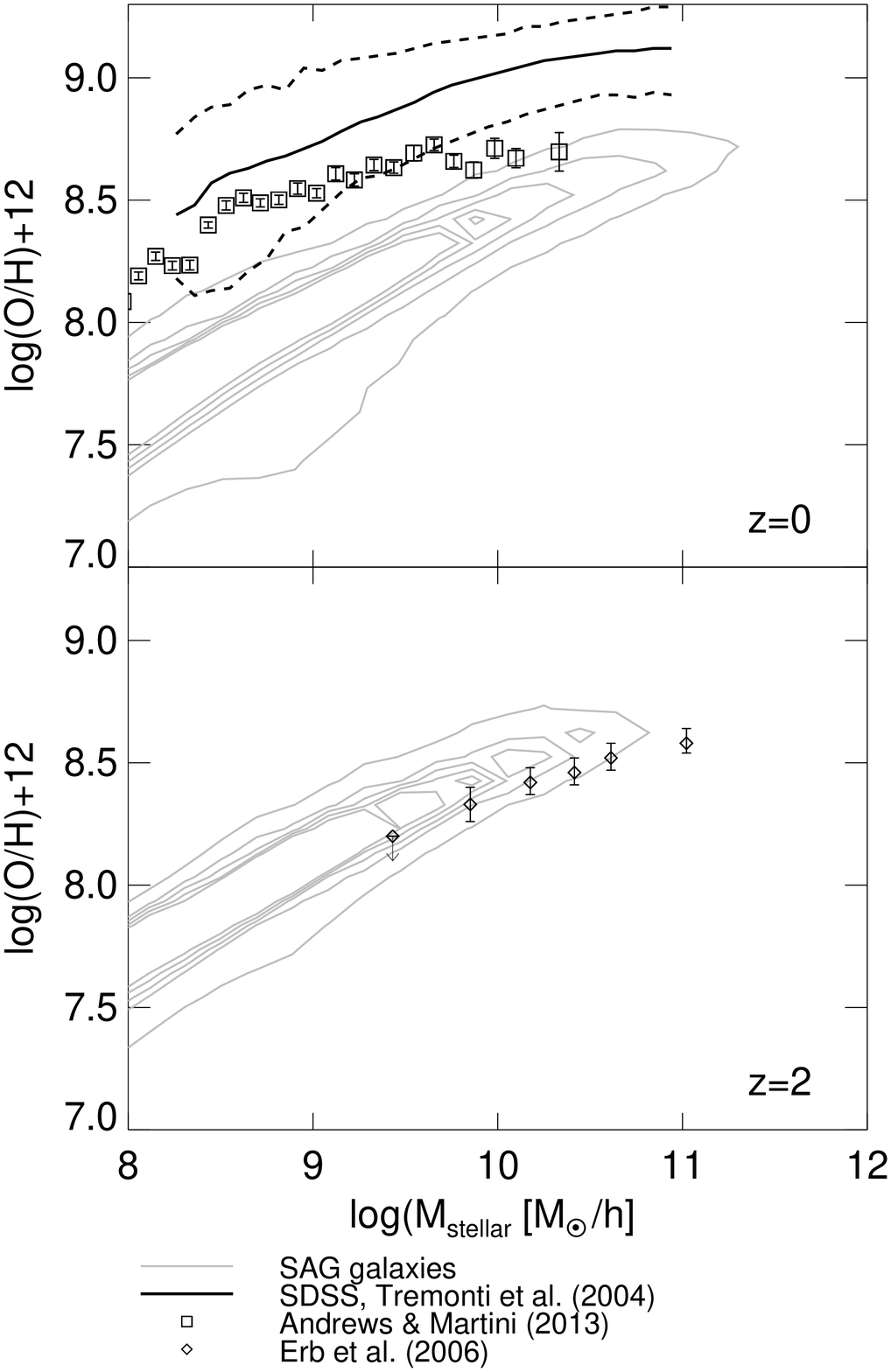}
\caption{Mass-metallicity relation predicted by \textsc{SAG} (contours). At $z=0$ (top panel) the model distribution is compared to the median value measured by \citet{Tremonti2004} (black solid line, with dashed lines as the 68 percentile) and \citet{Andrews2013} (black squares) using SDSS star-forming galaxies, while at $z=2$ (bottom panel) is compared to data from \citet{Erb2006} using UV-selected star-forming galaxies.}
\label{sagmassmet}
\end{center}
\end{figure}

Fig. \ref{sagsfrvol} shows the evolution of the cosmic SFR density, compared to observational data compiled by \citet{Behroozi2013} scaled to our adopted IMF. Our model predicts consistently the rise and fall of the cosmic SFR going from higher to lower redshifts, although at $z=0$ we have some excess compared to observations. Compared to \citet{Magnelli2013} data based on \textit{Herschel} observations (corrected to a Salpeter IMF as well), we underpredict the cosmic SFR by a factor of 1.2 at $z=0$, and by a factor of 3 in the range $z=1-2$. Note that this does not minimize the potential of the count matching technique, since when performing it we are only concerned about reproducing the observed trend for the cosmic SFR across redshift (not the exact values) which is successfully achieved with the model used.

Fig. \ref{sagmassmet} shows the predicted mass-metallicity relation at $z=0$ and 2. Compared to \citet{Tremonti2004} and \citet{Andrews2013} data, our cold gas phase metallicities at $z=0$ are sistematically shifted downwards.

If we consider the wide spread in the fit to the mass-metallicity relation given by different metallicity calibrations \citep{Kewley2008}, the gas metallicities from our model can be considered acceptable. The low gas metallicity values that characterize $z=0$ model galaxies might be related to the low SFR in the redshift range $z=2-3$ as is evident from the marginal agreement with observational data depicted in Fig. \ref{sagsfrvol}; the number of core collapse supernovae, which are the main source of oxygen, is directly related with the SFR. Levels of star formation may increase by considering the effects of accretion of material with missaligned angular momenta on the gaseous disc from which stars are formed \citep{Padilla2014}. On the other hand, feedback from supernova explosions plays a crucial role in determining the mass-metallicity relation. Both these processes are far from being well understood and their modelization needs improvements.

\begin{figure}
\begin{center}
\includegraphics[width=0.42\textwidth]{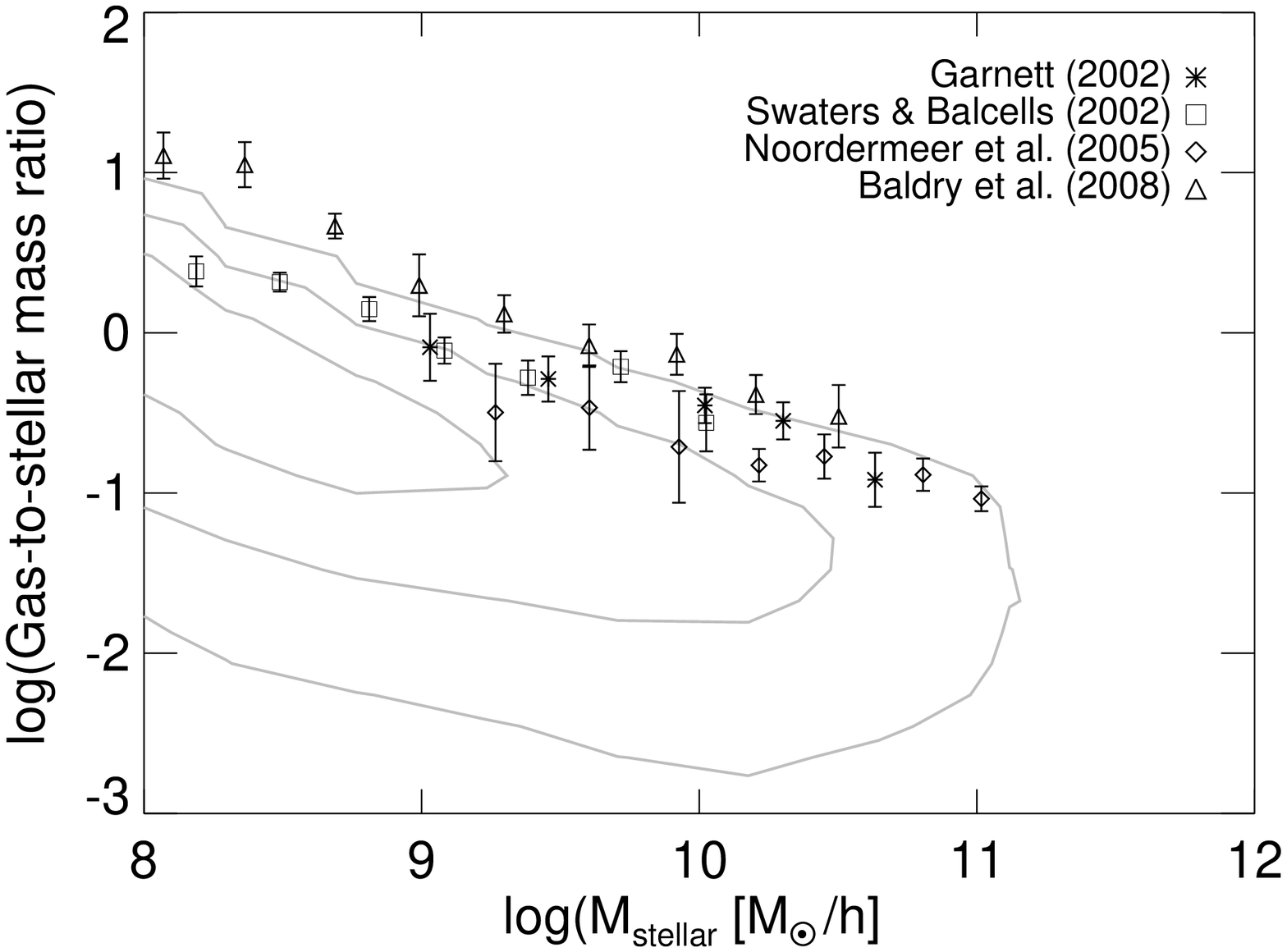}
\caption{Cold gas to stellar mass ratio as a function of stellar mass. Contours: predictions from \textsc{SAG}. Symbols: measured values by \citet{Garnett2002} (asterisks), \citet{Swaters2002} (open squares), \citet{Noordermeer2005} (open diamonds) and \citet{Baldry2008} (open triangles).}
\label{saggasfrac}
\end{center}
\end{figure}

Considering galaxies at $z=2$, we predict slightly higher cold gas metallicities  when compared to \citet{Erb2006} data. In any case, when applying the count matching technique we are interested only in the trends followed among properties, and not in the exact values; in this sense, the observed trend between mass and metallicity is well reproduced. This also applies to the ratio between cold gas and stellar mass as a function of stellar mass at $z=0$, shown in Fig. \ref{saggasfrac}. In our model, this ratio follows the observed trend from measurements by \citet{Garnett2002}, \citet{Swaters2002}, \citet{Noordermeer2005} and \citet{Baldry2008}, but underpredicts it by a factor of $\sim3$.

\section{Model for Bulge Sizes}\label{appbulges}

\begin{figure}
\begin{center}
\includegraphics[width=0.36\textwidth]{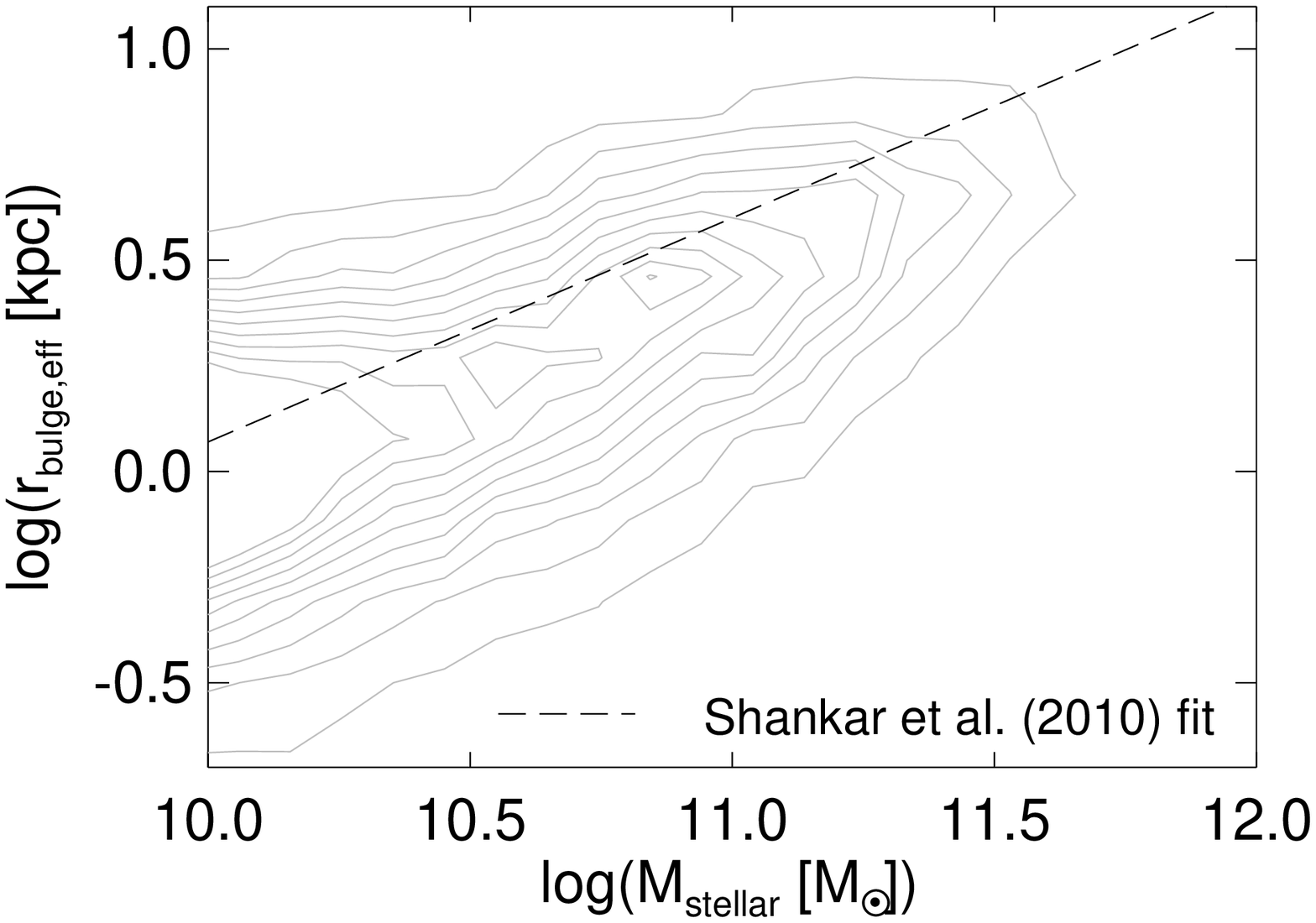}
\caption{Size-stellar mass relation for elliptical galaxies at $z=0$ in the \textsc{SAG} model (contours) and for a sample of SDSS ellipticals by \citet{Shankar2010} (linear fit, dashed black line).}
\label{remstellar}
\end{center}
\end{figure}

We implement a model for bulge sizes in \textsc{SAG}, based on the one introduced by \citet{Cole2000} with further improvements by \citet{Gonzalez2009}. In this approach, virial equilibrium and energy conservation are considered for computing the bulge size of a given galaxy.

Following \citet{Gonzalez2009}, the bulge size is computed when galaxy mergers and disc instabilities occur. When there is a merger, the bulge radius of the remnant galaxy $r_{\textmd{new,merger}}$ is given by
\begin{equation}
\frac{(M_1+M_2)^2}{r_{\textmd{new,merger}}}=\frac{M_1^2}{r_1}+\frac{M_2^2}{r_2}+\frac{f_{\textmd{orbit}}}{\bar{c}}\frac{M_1\,M_2}{r_1+r_2}, \label{}
\end{equation}
\noindent where the subscripts 1 and 2 stand for the central and satellite galaxies, respectively. In the case of a major merger, the masses involved are $M_i=M_{i,\textmd{stellar}}+M_{i,\textmd{cold gas}}+f_{\textmd{dark}}M_{i,\textmd{DM}}$ (with $i=1,\,2$), while for minor mergers are $M_1=M_{1,\textmd{stellar,bulge}}+f_{\textmd{dark}}M_{1,\textmd{DM}}$ and $M_2=M_{2,\textmd{stellar}}+f_{\textmd{dark}}M_{2,\textmd{DM}}$. Note that in all cases there is a contribution from the dark matter halo mass $M_{\textmd{DM}}$, determined by a factor with fiducial value $f_{\textmd{dark}}=2$. The form factor adopted is $\bar{c}=0.5$. The sizes involved are radii weighted by mass, i.e., $r_i=(M_{i,\textmd{bulge}}r_{i,\textmd{bulge}}+M_{i,\textmd{disc}}r_{i,\textmd{disc}})/(M_{i,\textmd{bulge}}+M_{i,\textmd{disc}})$ (with $i=1,\,2$).

When there is a disc instability, the bulge size $r_{\textmd{new,instab}}$ is given by
\begin{eqnarray}
\frac{c_B(M_{\textmd{bulge}}+M_{\textmd{disc}})^2}{r_{\textmd{new,instab}}}&=&\frac{c_B M_{\textmd{bulge}}^2}{r_{\textmd{bulge}}}+\frac{c_D M_{\textmd{disc}}^2}{r_{\textmd{disc}}}\nonumber\\&&+f_{\textmd{int}}\frac{M_{\textmd{bulge}}M_{\textmd{disc}}}{r_{\textmd{bulge}}+r_{\textmd{disc}}}, \label{}
\end{eqnarray}
\noindent where the disc mass includes the contribution of both the stellar and cold gas components. As in \citet{Cole2000}, the adopted factors are $c_D=0.49$, $c_B=0.45$ and $f_{\textmd{int}}=2$.

In this work we introduce an update of the bulge size when a starburst occurs, inspired by the model for disc instabilities. In this case the bulge size immediately after the starburst event $r_{\textmd{new,starb}}$ is given by
\begin{eqnarray}
\frac{c_B(M_{\textmd{bulge}}+M_{\textmd{starb}})^2}{\Delta r_{\textmd{new,starb}}}&=&\frac{c_B M_{\textmd{bulge}}^2}{r_{\textmd{bulge}}}+\frac{c_D M_{\textmd{starb}}^2}{r_{\textmd{starb}}}\nonumber\\&&+f_{\textmd{int}}\frac{M_{\textmd{bulge}}M_{\textmd{starb}}}{r_{\textmd{bulge}}+r_{\textmd{starb}}},\\r_{\textmd{new,starb}}&=&r_{\textmd{bulge}}+\Delta r_{\textmd{new,starb}},\label{}
\end{eqnarray}
\noindent where $M_{\textmd{starb}}$ is the mass of stars formed in the starburst and $r_{\textmd{starb}}$ is a characteristic scale length associated with it. For simplicity, we take $r_{\textmd{starb}}=r_{\textmd{disc}}$. The adopted factors $c_D$, $c_B$ and $f_{\textmd{int}}=2$ are the same as for disc instabilities.

For validating this model in \textsc{SAG}, we present our predicted size-stellar mass relation for elliptical galaxies at $z=0$ in Fig. \ref{remstellar}. The size is characterized by the bulge effective radius; the projected density is assumed to be well described by a de Vaucouleurs profile, $\rho(r)\propto r^{1/4}$. Our model is in good agreement with the linear fit obtained by \citet{Shankar2010} for a sample of ellipticals in SDSS.

\label{lastpage}


\begin{thebibliography}{99}

\bibitem[{Andrews} \& {Martini}(2013)]{Andrews2013}
Andrews, B. H. \& Martini, P., Astrophys. J. 765 (2013), 140.

\bibitem[{Baldry} {et~al.}(2008)]{Baldry2008}
Baldry, I. K. et al., Mon. Not. R. Astron. Soc. 388 (2008), 945.

\bibitem[{Barger} {et~al.}(1998)]{Barger1998}
Barger, A. J. et al., Nature 394 (1998), 248.

\bibitem[{Baugh} {et~al.}(2005)]{Baugh2005}
Baugh, C. M. et al., Mon. Not. R. Astron. Soc. 356 (2005), 1191.

\bibitem[{Behroozi} {et~al.}(2010)]{Behroozi2010}
Behroozi, P. S. et al., Astrophys. J. 717 (2010), 379.

\bibitem[{Behroozi} {et~al.}(2013)]{Behroozi2013}
Behroozi, P. S. et al., Astrophys. J. 770 (2013), 57.

\bibitem[{Bertin} \& {Arnouts}(1996)]{Bertin1996}
Bertin, E. \& Arnouts, S., Astron. \& Astrophys. Suppl. Ser. 117 (1996), 393.

\bibitem[{B\'ethermin} {et~al.}(2012)]{Bethermin2012}
B\'ethermin, M. et al., Astrophys. J. Letters 757 (2012), L23.

\bibitem[{Bhattacharjee} {et~al.}(2014)]{Bhattacharjee2014}
Bhattacharjee, P. et al., Astrophys. J. 785 (2014), 63.

\bibitem[{Blain}(1999)]{Blain1999}
Blain, A. W., Mon. Not. R. Astron. Soc. 304 (1999), 669.

\bibitem[{Blain} {et~al.}(1999)]{Blain1999b}
Blain, A. W. et al., Mon. Not. R. Astron. Soc. 302 (1999), 632.

\bibitem[{Blain} {et~al.}(2002)]{Blain2002}
Blain, A. W. et al., Phys. Rep. 369 (2002), 111.

\bibitem[{Borys} {et~al.}(2005)]{Borys2005}
Borys, C. et al., Astrophys. J. 635 (2005), 853.

\bibitem[{Bruzual} \& {Charlot}(2003)]{Bruzual2003}
Bruzual, G. \& Charlot, S., Mon. Not. R. Astron. Soc. 344 (2003), 1000.

\bibitem[{Chen} {et~al.}(2013)]{Chen2013}
Chen, C. C. et al., Astrophys. J. 776 (2013), 131.

\bibitem[{Chen} {et~al.}(2014)]{Chen2014}
Chen, C. C. et al., Astrophys. J. 789 (2014), 12.

\bibitem[{Chapman} {et~al.}(2004)]{Chapman2004}
Chapman, S. C. et al., Astrophys. J. 611 (2004), 732.

\bibitem[{Chapman} {et~al.}(2005)]{Chapman2005}
Chapman, S. C. et al., Astrophys. J. 622 (2005), 772.

\bibitem[{Cole} {et~al.}(2000)]{Cole2000}
Cole, S. et al., Mon. Not. R. Astron. Soc. 319 (2000), 168.

\bibitem[{Conroy} {et~al.}(2006)]{Conroy2006}
Conroy, C. et al., Astrophys. J. 647 (2006), 201.

\bibitem[{Coppin} {et~al.}(2006)]{Coppin2006}
Coppin, K. et al., Mon. Not. R. Astron. Soc. 372 (2006), 1621.

\bibitem[{Cora}(2006)]{Cora2006}
Cora, S. A., Mon. Not. R. Astron. Soc. 368 (2006), 1540.

\bibitem[{Cowley} {et~al.}(2014)]{Cowley2014}
Cowley, W. I. et al., arXiv:1406.0855 (2014).

\bibitem[{Croton} {et~al.}(2006)]{Croton2006}
Croton, D. J. et al., Mon. Not. R. Astron. Soc. 365 (2006), 11.

\bibitem[{Dav\'e} {et~al.}(2014)]{Dave2014}
Dav\'e, R. et al., Mon. Not. R. Astron. Soc. 404 (2014), 1355.

\bibitem[{De Lucia} {et~al.}(2004)]{DeLucia2004}
De Lucia, G. et al., Mon. Not. R. Astron. Soc. 349 (2004), 1101.

\bibitem[{Dekel} {et~al.}(2009)]{Dekel2009}
Dekel, A. et al., Nature 457 (2009), 451.

\bibitem[{Er} {et~al.}(2013)]{Er2013}
Er, X. et al., Mon. Not. R. Astron. Soc. 430 (2013), 1423.

\bibitem[{Erb} {et~al.}(2006)]{Erb2006}
Erb, D. K. et al., Astrophys. J. 644 (2006), 813.

\bibitem[{Flynn} {et~al.}(2006)]{Flynn2006}
Flynn, C. et al., Mon. Not. R. Astron. Soc. 372 (2006), 1149.

\bibitem[{Fontanot} {et~al.}(2007)]{Fontanot2007}
Fontanot, F. et al., Mon. Not. R. Astron. Soc. 382 (2007), 903.

\bibitem[{Gargiulo} {et~al.}(2014)]{Gargiulo2014}
Gargiulo, I. D. et al., arXiv:1402.3296 (2014).

\bibitem[{Garnett}(2002)]{Garnett2002}
Garnett, D. R., Astrophys. J. 581 (2002), 1019.

\bibitem[{Gonz\'alez} {et~al.}(2009)]{Gonzalez2009}
Gonz\'alez, J. E. et al., Mon. Not. R. Astron. Soc. 397 (2009), 1254.

\bibitem[{Graham} {et~al.}(2005)]{Graham2005}
Graham, A. W. et al., Astron. J. 130 (2005), 1535.

\bibitem[{Granato} {et~al.}(2000)]{Granato2000}
Granato, G. L. et al., Astrophys. J. 542 (2000), 710.

\bibitem[{Gruppioni} {et~al.}(2013)]{Gruppioni2013}
Gruppioni, C. et al., Mon. Not. R. Astron. Soc. 432 (2013), 23.

\bibitem[{Hatsukade} {et~al.}(2013)]{Hatsukade2013}
Hatsukade, B. et al., Astrophys. J. Letters 769 (2013), L27.

\bibitem[{Hayward} {et~al.}(2011)]{Hayward2011}
Hayward, C. C. et al., Astrophys. J. 743 (2011), 159.

\bibitem[{Hayward} {et~al.}(2013)]{Hayward2013}
Hayward, C. C. et al., Mon. Not. R. Astron. Soc. 428 (2013), 2529.

\bibitem[{Hayward} {et~al.}(2013b)]{Hayward2013b}
Hayward, C. C. et al., Mon. Not. R. Astron. Soc. 434 (2013), 2572.

\bibitem[{Hickox} {et~al.}(2012)]{Hickox2012}
Hickox, R. C. et al., Mon. Not. R. Astron. Soc. 421 (2012), 284.

\bibitem[{Hodge} {et~al.}(2013)]{Hodge2013}
Hodge, J. A. et al., Mon. Not. R. Astron. Soc. 768 (2013), 91.

\bibitem[{Hogg} {et~al.}(2002)]{Hogg2002}
Hogg, D. W. et al., arXiv:astro-ph/0210394 (2002).

\bibitem[{Holland} {et~al.}(1999)]{Holland1999}
Holland, W. S. et al., Mon. Not. R. Astron. Soc. 303 (1999), 659.

\bibitem[{Hughes} {et~al.}(1998)]{Hughes1998}
Hughes, D. H. et al., Nature 394 (1998), 241.

\bibitem[{Jarosik} {et~al.}(2011)]{Jarosik2011}
Jarosik, N. et al., Astrophys. J. Suppl. Ser. 192 (2011), 14.

\bibitem[{Jonsson}(2006)]{Jonsson2006}
Jonsson, P., Mon. Not. R. Astron. Soc. 372 (2006), 2.

\bibitem[{Karim} {et~al.}(2013)]{Karim2013}
Karim, A. et al., Mon. Not. R. Astron. Soc. 432 (2013), 2.

\bibitem[{Kennicutt}(1983)]{Kennicutt1983}
Kennicutt, R. C., Astrophys. J. 272 (1983), 54.

\bibitem[{Kewley} \& {Ellison}(2008)]{Kewley2008}
Kewley, L. J. \& Ellison, S. L., Astrophys. J. 681 (2008), 1183.

\bibitem[{Lagos, Cora \& Padilla}(2008)]{Lagos2008}
Lagos, C. del P., Cora, S. A. \& Padilla, N. D., Mon. Not. R. Astron. Soc. 388 (2008), 587.

\bibitem[{Laird} {et~al.}(2010)]{Laird2010}
Laird, E. S. et al., Mon. Not. R. Astron. Soc. 401 (2010), 2763.

\bibitem[{Lima} {et~al.}(2010)]{Lima2010}
Lima, M. et al., Astrophys. J. Letters 717 (2010), L31.

\bibitem[{Magdis} {et~al.}(2012)]{Magdis2012}
Magdis, G. E. et al., Astrophys. J. 760 (2012), 6.

\bibitem[{Magnelli} {et~al.}(2013)]{Magnelli2013}
Magnelli, B. et al., Astron. \& Astrophys. 553 (2013), A132.

\bibitem[{Micha\l{}owski} {et~al.}(2010)]{Michalowski2010}
Micha\l{}owski, M. et al., Astron. \& Astrophys. 514 (2010), A67.

\bibitem[{Moster} {et~al.}(2010)]{Moster2010}
Moster, B. P. et al., Astrophys. J. 710 (2010), 903.

\bibitem[{Murray} \& {Rahman}(2010)]{Murray2010}
Murray, N. \& Rahman, M., Astrophys. J. 709 (2010), 424.

\bibitem[{Noll} {et~al.}(2009)]{Noll2009}
Noll, S. et al., Astron. \& Astrophys. 507 (2009), 1793.

\bibitem[{Noordermeer} {et~al.}(2005)]{Noordermeer2005}
Noordermeer, E. et al., Astron. \& Astrophys. 442 (2005), 137.

\bibitem[{Padilla} {et~al.}(2014)]{Padilla2014}
Padilla, N. D. et al., Mon. Not. R. Astron. Soc. 443 (2014), 2801.

\bibitem[{Ruiz} {et~al.}(2014)]{Ruiz2014}
Ruiz, A. N. et al., arXiv:1310.7034 (2014).

\bibitem[{Rujopakarn} {et~al.}(2011)]{Rujopakarn2011}
Rujopakarn, W. et al., Astrophys. J. 726 (2011), 93.

\bibitem[{Salpeter}(1955)]{Salpeter1955}
Salpeter, E. E., Astrophys. J. 121 (1955), 161.

\bibitem[{Sanders} \& {Mirabel}(1996)]{Sanders1996}
Sanders, D. B. \& Mirabel, I. F., Annu. Rev. Astron. Astrophys 34 (1996), 749.

\bibitem[{Scott} {et~al.}(2008)]{Scott2008}
Scott, K. S. et al., Mon. Not. R. Astron. Soc. 385 (2008), 2225.

\bibitem[{Shankar} {et~al.}(2010)]{Shankar2010}
Shankar, F. et al., Mon. Not. R. Astron. Soc. 403 (2010), 117.

\bibitem[{Silva} {et~al.}(1998)]{Silva1998}
Silva, L. et al., Astrophys. J. 509 (1998), 103.

\bibitem[{Simpson} {et~al.}(2014)]{Simpson2014}
Simpson, J. M et al., Astrophys. J. 788 (2014), 125.

\bibitem[{Siringo} {et~al.}(2009)]{Siringo2009}
Siringo, G. et al., Astron. \& Astrophys. 497 (2009), 945.

\bibitem[{Smail} {et~al.}(2004)]{Smail2004}
Smail, I. et al., Astrophys. J. 616 (2004), 71.

\bibitem[{Smol\v{c}i\'c} {et~al.}(2012)]{Smolcic2012}
Smol\v{c}i\'c, V. et al., Astron. \& Astrophys. 548 (2012), A4.

\bibitem[{Somerville} {et~al.}(2012)]{Somerville2012}
Somerville, R. S. et al., Mon. Not. R. Astron. Soc. 423 (2012), 1992.

\bibitem[{Springel}(2005)]{Springel2005}
Springel, V., Mon. Not. R. Astron. Soc. 364 (2005), 1105.

\bibitem[{Swaters} \& {Balcells}(2002)]{Swaters2002}
Swaters, R. A. \& Balcells, M., Astron. \& Astrophys. 390 (2002), 863.

\bibitem[{Swinbank} {et~al.}(2004)]{Swinbank2004}
Swinbank, A. M. et al., Astrophys. J. 617 (2004), 64.

\bibitem[{Swinbank} {et~al.}(2006)]{Swinbank2006}
Swinbank, A. M. et al., Mon. Not. R. Astron. Soc. 371 (2006), 465.

\bibitem[{Swinbank} {et~al.}(2008)]{Swinbank2008}
Swinbank, A. M. et al., Mon. Not. R. Astron. Soc. 391 (2008), 420.

\bibitem[{Swinbank} {et~al.}(2014)]{Swinbank2014}
Swinbank, A. M. et al., Mon. Not. R. Astron. Soc. 438 (2014), 1267.

\bibitem[{Tacconi} {et~al.}(2006)]{Tacconi2006}
Tacconi, L. J. et al., Astrophys. J. 640 (2006), 228.

\bibitem[{Tacconi} {et~al.}(2008)]{Tacconi2008}
Tacconi, L. J. et al., Astrophys. J. 680 (2008), 246.

\bibitem[{Tacconi} {et~al.}(2010)]{Tacconi2010}
Tacconi, L. J. et al., Nature 463 (2010), 781.

\bibitem[{Targett} {et~al.}(2013)]{Targett2013}
Targett, T. A. et al., Mon. Not. R. Astron. Soc. 432 (2013), 2012.

\bibitem[{Tecce} {et~al.}(2010)]{Tecce2010}
Tecce, T. E. et al., Mon. Not. R. Astron. Soc. 408 (2010), 2008.

\bibitem[{Tremonti} {et~al.}(2004)]{Tremonti2004}
Tremonti, C. A. et al., Astrophys. J. 613 (2004), 898.

\bibitem[{Wang} {et~al.}(2013)]{Wang2013}
Wang, S. X. et al., Astrophys. J. 778 (2013), 179

\bibitem[{Wardlow} {et~al.}(2011)]{Wardlow2011}
Wardlow, J. L., Mon. Not. R. Astron. Soc. 415 (2011), 1479.

\bibitem[{Weiss} {et~al.}(2009)]{Weiss2009}
Weiss, A. et al., Astrophys. J. 707 (2009), 1201.

\bibitem[{Weiss} {et~al.}(2013)]{Weiss2013}
Weiss, A. et al., Astrophys. J. 767 (2013), 88.

\end{thebibliography}
\end{document}